\documentclass[12pt]{article}
\usepackage{natbib}
\usepackage{graphicx}
\usepackage{amsmath,amsfonts,amsthm}
\newcommand{\noi}{\noindent}

\usepackage{bm}
\usepackage{amsfonts}

\def\sqr#1#2{{\vcenter{\hrule height.#2pt
\hbox{\vrule width.#2pt height #1pt \kern#1pt \vrule width.#2pt}
\hrule height .#2pt}}}

\title{Robust Generalised Quadratic Discriminant Analysis}
\author{Abhik Ghosh$^{1*}$\\
 Rita SahaRay $^1$\\ Sayan  Chakrabarty$^2$\\Sayan Bhadra$^1$}

\begin{document}
\date{}
%\normalsize
\maketitle
\begin{center}\noindent
\\$^1$Interdisciplinary Statistical Research Unit,\\ Indian Statistical Institute,  Kolkata-700108, India.\\
$^2$ Department of Statistics,\\University of Illinois at Urbana-Champaign, Illinois 61820, USA.
\end{center}
\noindent \begin{abstract} \noindent 
Quadratic discriminant analysis (QDA) is a widely used  statistical tool to classify observations from different multivariate Normal populations. 
The generalized quadratic discriminant analysis (GQDA) classification rule/classifier,  which generalizes the QDA and  the minimum Mahalanobis distance (MMD) classifiers 
to discriminate between populations with underlying elliptically symmetric distributions competes quite favorably with the  QDA classifier when it is optimal 
and performs much better when QDA fails under non-Normal  underlying distributions, e.g. Cauchy distribution.  
However, the  classification rule  in GQDA is based on the sample  mean vector  and the sample  dispersion  matrix of a  training sample,
which are extremely non-robust under data contamination. 
In real world, since it is  quite common to face data  highly vulnerable to outliers, 
the lack of robustness of the classical estimators of the mean vector and the dispersion  matrix reduces 
the  efficiency of the  GQDA classifier significantly,   increasing  the misclassification errors.   
%The study on the effectiveness of various   robust methods for linear and quadratic discriminant analysis  has attracted  many researchers through  recent past decades. 
The present paper investigates the performance of the GQDA classifier  when the classical estimators 
of the mean vector and the dispersion  matrix used therein  are replaced by various  robust counterparts. 
Applications to various real data sets as well as simulation  studies reveal far better performance of the proposed robust versions of the GQDA classifier. 
A Comparative study has been made to advocate the  appropriate choice  of the robust estimators 
to be used in a specific situation of the  degree of contamination of the data sets. 
\end{abstract}
\thispagestyle{empty}

\vspace*{\fill}

\noindent \hrule
\vspace{.2cm}
%Stat-Math Technical Report no:4/2002  \hfill 09.07.02
%Stat-Math Technical Report no:4/2002  \hfill 23.05.06

\noi *Corresponding author. Fax: 91-33-2577-3104. Email: abhik.ghosh@isical.ac.in.

\begin{description}
%\item[AMS Subject Classification (2010) : ]
\item[Keywords and Phrases :]
Linear discriminant analysis, Quadratic discriminant analysis,
Generalized quadratic discriminant analysis, Robust estimators 

\end{description}

\pagebreak

\section{Introduction}
\setcounter{equation}{0}

Discriminant analysis is a very widely used statistical tool  to assign an 
individual to any of the  $k (\geq 2)$ populations  on the basis of a $p$ dimensional feature vector.  
Usually it is assumed that the underlying distribution of the feature vectors is multivariate Normal. 
Under this usual assumption, with equal  dispersion   matrices of different underlying populations, 
 linear discriminant analysis (LDA) leads to a classification rule based on minimizing 
the Mahalanobis distance of the  new observation from the  mean vector  of a particular underlying population in question \citep{Zollanvari/etc:2013}. 
This classification rule is also referred to as minimum Mahalanobis distance (MMD) classification rule. 
When the assumption of equality of the  dispersion   matrices of the underlying populations  is not tenable, 
quadratic discriminant analysis (QDA) is used, where the classification rule involves the ratio of the determinants of 
the  dispersion    matrices, apart from the  Mahalanobis distances. 
More recently, several methodological and computational advancements of LDA and QDA have been developed 
to generate faster and better performances under specific problems; see, e.g., 
\cite{Hua/etc:2005,Wang/etc:2008,Park/Park:2008,Na/etc:2010,Suzuki/Itoh:2010,Daqi/etc:2014,Ye/etc:2017}, among many others.

It  is  worthwhile to note that, with underlying elliptically symmetric distributions which are not necessarily Normal, 
the classification rule  to discriminate between these populations also involve a similar factor as QDA. 
Noting this interesting phenomenon and observing that QDA does not perform well in discriminating populations with non-Normal distributions, 
\cite{Bose/etc:2015} generalized the QDA and MMD classifiers. 
Their proposed method, termed as generalized quadratic discriminant analysis (GQDA), is a simple  nonparametric method 
which is adaptive to any given data set by choosing a threshold value to  make the decision on where to classify the new observation. 
The performance of the classifier under this flexible method matches with that of the QDA classifier when it is optimal 
and compares quite favorably with the  other established complex nonparametric classifiers. 

However, like QDA, the proposed GQDA is also based on the  mean  vectors and the dispersion    matrices of the populations, 
which being unknown, need to be estimated from  a part of the data, namely the training   set.  
In practice, quite often  the populations get contaminated because of the presence of outlying observations. 
The susceptibility of the classical sample  mean vector   and the sample  dispersion  matrix  to outlying observations present in the training   set  
tends to misclassify the new observation, leading to unreliability of the classical LDA and QDA as well as  GQDA. 
As the simple cost effective classifier in  GQDA has competitive and sometimes even better performance in  a wide range of distributions, 
not only  limited to the   Normal distribution, it is worthwhile to  make  this procedure  robust in the face of contamination. 
This necessity motivates the present authors to make a thorough investigation of the performance of  the GQDA classifier 
in the presence of contamination  and undertake a comparative study of different robust versions  of GQDA,
replacing the classical estimators in the GQDA classifier by their robust counterparts.

The issue of non-robustness of LDA and QDA has been addressed by several researchers 
replacing the classical estimators in LDA and QDA  by their robust counterparts.   
\cite{Randles/etc:1978}  proposed to use  M estimators and used a rank based rule to estimate the threshold value for discrimination between two populations.  
\cite{Todorov/etc:1990,Todorov/etc:1994} worked with  minimum covariance determinant (MCD) estimators, 
while  \cite{Chork/Rousseeuw:1992} and  \cite{Kim/etc:2006} used minimum volume ellipsoid (MVE) estimators.  
\cite{Croux/Dehon:2001} advocated S-estimators while  \cite{Hubert/etc:2012} applied MCD estimates computed by the FAST MCD algorithm.  
In the present article, we focus on the development of the robust  generalized quadratic discriminant analysis (RGQDA) 
for two class as well as multi-class (more than two classes)  classification problems,
using such robust estimators of  the mean vector and the dispersion  matrix along with a detailed (empirical) comparative study.
Based on our investigations, suggestions  are also made on the  specific choice of the robust estimators  
under situations with different degree of contamination.

The paper is organized as follows. 
Section  2 gives a brief overview of  GQDA  and illustrate its unreliability in the presence of outliers 
through a simulated example of Normal distribution. For ready reference,  
different robust estimators of the   mean vector   and   the dispersion   matrix available in the literature are  briefly reviewed  in Section 3.  
Simulation studies related to two class as well as multi class classifications are presented in Section 4 
to illustrate  the improvement of the  performance of our proposed RGQDA   over   GQDA. 
The potential of  RGQDA is compared   with GQDA  using some real data sets in Section 5.   
Finally, Section 6 gives the concluding remarks and  future directions of further research.

\section{The GQDA classifier and its non-robustness}
\label{SEC:GQDA}
\setcounter{equation}{0}

In this section,  we first present a brief overview of GQDA proposed by \cite{Bose/etc:2015} 
for two-class as well as multi-class classification problem.

\subsection{Two-class classification}

Let us first consider the simpler case where  an object is classified into one of the two  competing populations  or classes
using a decision rule formed on the basis of an observation  $\bm{x}=(x_1,\ldots,x_p)'$ 
on a $p$-variate random  feature vector $\bm{X}$.  
The rule is devised using a sample  of $n$ observations on $\bm{X}$, called the training  set,  
which represents observations from both the populations. 
It is assumed that, for the  $j$th population, $j=1,2$, 
the random vector $\bm{X}$ has a probability density function $f_j(\bm{x})$,   
and $ \pi_j$ is the \textit{a priori} probability for an observation to belong to this population, where $ \pi_1 + \pi_2 =1$. 
Whenever the assumption of equality of prior probabilities $\pi_1$ and $\pi_2$ holds,  the optimal Bayes rule assigns the 
observation $\bm{x}$ to that population $j$, for which the density function evaluated at $\bm{x}$ is larger. 
Thus the  feature space $\cal X$ is  essentially partitioned into $R_{1}$ and $R_2$  such that  
the  object is classified into  population 1 or population 2 depending on   whether $\bm{x} \in R_1 $ or $\bm{x} \in R_2 $,  respectively, 
where
$$R_{1}=\left \{\bm{x}:\frac{f_{1}(\bm{x})}{f_{2}(\bm{x})}\geq 1\right \}=\left \{\bm{x}:\log\frac{f_{1}(\bm{x})}{f_{2}(\bm{x})}\geq 0\right \},$$
$$R_{2}=\left \{\bm{x}:\frac{f_{1}(\bm{x})}{f_{2}(\bm{x})}<1\right \}=\left \{\bm{x}:\log\frac{f_{1}(\bm{x})}{f_{2}(\bm{x})}< 0\right \}.$$

In the case of  two underlying populations having   multivariate Normal distribution with the mean  vectors  $\bm{\mu_{1}}$, $\bm{\mu_{2}}$ and the dispersion   matrices $\bm{\Sigma}_1$ and $\bm{\Sigma}_2$,  respectively, we get 
\begin{eqnarray}
\log\frac{f_{1}(\bm{x})}{f_{2}(\bm{x})}
%&=&\frac{1}{2}\!\log\left(\frac{|\bm{\Sigma}_2|}{|\bm{\Sigma}_1|}\right)+\frac{1}{2}[(\bm{x}-\bm{\mu_{2}})'\bm{\Sigma}_{2}^{-1}                                                  (\bm{x}-\bm{\mu_{2}}) \nonumber 
%  -(\bm{x}-\bm{\mu_{1}})'\bm{\Sigma}_{1}^{-1}(\bm{x}-\bm{\mu_{1}})] \nonumber \\
&=&\frac{1}{2}\!\log\left(\frac{\left|\bm{\Sigma}_{2}\right|}{\left|\bm{\Sigma}_{1}\right|}\right)+\frac{1}{2}{\Delta_d}^2,\mbox{ say},
\end{eqnarray}
where
${\Delta_d}^2=\left[(\bm{x}-\bm{\mu_{2}})'\bm{\Sigma}_{2}^{-1}(\bm{x}-\bm{\mu_{2}})
-(\bm{x}-\bm{\mu_{1}})'\bm{\Sigma}_{1}^{-1}(\bm{x}-\bm{\mu_{1}})\right]$
is nothing but the difference of the squared Mahalanobis distances of $\bm{x}$ from the two  populations. 
Thus, for the QDA classification rule, we have
\begin{eqnarray}
\label{eq1}
R_{1} &=&\left \{\bm{x}: \frac{1}{2}\log\left(\frac{\left|\bm{\Sigma}_{2}\right|}{\left|\bm{\Sigma}_{1}\right|}\right)+\frac{1}{2}{\Delta_d}^2\geq 0\right \}
=\left \{\bm{x}:{\Delta_d}^2 \geq \log\left(\frac{\left|\bm{\Sigma}_{1}\right|}{\left|\bm{\Sigma}_{2}\right|}\right)\right \},
\nonumber \\
R_{2} & = & \mathcal{X} - R_{1} = 
\left \{\bm{x}:{\Delta_d}^2 < \log\left(\frac{\left|\bm{\Sigma}_{1}\right|}{\left|\bm{\Sigma}_{2}\right|}\right)\right \}.
\end{eqnarray}
It is to be noted  that when the assumption of equality of  $\bm{\Sigma}_1$  and  $\bm{\Sigma}_2$, holds, 
simplifying  (\ref{eq1}) we get  $R_1=\left \{\bm{x}:{\Delta_d}^2 \geq 0 \right \}$, turning the QDA  rule  identical to the MMD rule. 
However, in  the cases where this  assumption of the equality  of  the dispersion   matrices does not hold,  
the MMD rule  fails and the QDA rule turns out to be optimal. 
On the other hand,  it is  quite likely  that the MMD rule will have a better performance than the QDA rule 
in terms of reducing the misclassification error, if the underlying population probability densities are not Normal.

\noi 
Now we consider two underlying populations  having $p$-variate $t$-distribution with $q$ degrees of freedom (d.f.) with densities
$$
f_{j}(\bm{x})=A\:|\bm{\Sigma}_j|^{-\frac{1}{2}}\left[1+\frac{1}{q}(\bm{x}-\bm{\mu_{j}})'\bm{\Sigma}_{j}^{-1}(\bm{x}-\bm{\mu_{j}})\right]^{-\frac{p+q}{2}},
\ j=1,2,
$$                    
where 
$A ={\Gamma \left(\frac{p+q}{2}\right)}\:q^{p/2}\:\pi^{p/2}/{\Gamma \left(\frac{q}{2}\right)}$.
Thus,
\begin{equation}
\frac{f_{1}(\bm{x})}{f_{2}(\bm{x})}=\frac{\left|\bm{\Sigma}_{2}\right|^{1/2}}{\left|\bm{\Sigma}_{1}\right|^{1/2}}\;\left[\frac{1+\frac{1}{q}(\bm{x}-\bm{\mu_{2}})'\bm{\Sigma}_{2}^{-1}(\bm{x}-\bm{\mu_{2}})}{1+\frac{1}{q}(\bm{x}-\bm{\mu_{1}})'\bm{\Sigma}_{1}^{-1}(\bm{x}-\bm{\mu_{1}})}\right]^{\frac{p+q}{2}},
\nonumber
\end{equation}
and it follows that
\begin{eqnarray*}
\log\frac{f_{1}(\bm{x})}{f_{2}(\bm{x})}&=&\frac{1}{2}\log\frac{|\bm{\Sigma}_2|}{|\bm{\Sigma}_1|} \\
&&\\
&+&\frac{p+q}{2}\left[\log\{1+\frac{1}{q}(\bm{x}-\bm{\mu_2})'\bm{\Sigma}_{2}^{-1}(\bm{x}-\bm{\mu_2})\}-\log\{1+\frac{1}{q}(\bm{x}-\bm{\mu_1})'\bm{\Sigma}_{1}^{-1}(\bm{x}-\bm{\mu_1})\}\right]\\
&&\\
&\cong &\frac{1}{2}\log\left(\frac{|\bm{\Sigma}_{2}|}{|\bm{\Sigma}_{1}|}\right) + \frac{p+q}{2q}\left[(\bm{x}-\bm{\mu_2})'\bm{\Sigma}_{2}^{-1}
(\bm{x}-\bm{\mu_2}) -(\bm{x}-\bm{\mu_1)}'\bm{\Sigma}_{1}^{-1}(\bm{x}-\bm{\mu_1})\right]\\
&&\\
&=&\frac{1}{2}\log\left(\frac{|\bm{\Sigma}_{2}|}{|\bm{\Sigma}_{1}|}\right)+\left(\frac{1}{2}+\frac{p}{2q}\right){\Delta_d}^2,
\end{eqnarray*}
using the Taylor series expansion of terms of the type $\log(1+x)$ and neglecting higher-order terms, presuming $q$ to be sufficiently large.
Therefore, we get
\begin{eqnarray}
\label{eq2}
R_{1}&=&\left \{\bm{x}:\frac{1}{2}\log\left(\frac{\left|\bm{\Sigma}_{2}\right|}{\left|\bm{\Sigma}_{1}\right|}\right)+(\frac{1}{2}+\frac{p}{2q}){\Delta_d}^2\geq 0\right\}. \nonumber \\
&& \nonumber \\
& =&\left \{\bm{x}:{\Delta_d}^2 \geq \frac{q}{p+q}\log\left(\frac{\left|\bm{\Sigma}_{1}\right|}{\left|\bm{\Sigma}_{2}\right|}\right)\right\}.
\end{eqnarray}
It is known that as  $q \rightarrow \infty$, the $t$-distribution with $q$ degrees of freedom approaches the Normal distribution, hence  
  the rule (\ref{eq2}) boils down to the QDA rule in this case, as expected.
      
In fact it has been shown by \cite{Bose/etc:2015} that, for the class of elliptically symmetric distributions 
with the probability density function having  the form
$$f(\bm{x})=\frac{1}{|\bm{\Sigma}|^{\frac{1}{2}}}g((\bm{x}-\bm{\mu)}'\bm{\Sigma}^{-1}(\bm{x}-\bm{\mu})),
$$
the Bayes rule leads to the partition
\begin{equation}\label{ellip}
R_{1}=\left \{\bm{x}:\frac{1}{2}\log\left(\frac{\left|\bm{\Sigma}_{2}\right|}{\left|\bm{\Sigma}_{1}\right|}\right)+k\;{\Delta_d}^2\geq 0\right\},
\end{equation}
where $k$ may depend on $\bm{x}$.

Therefore, combining (\ref{eq1}), (\ref{eq2}),  and (\ref{ellip}) and 
denoting $\log\frac{|\bm{\Sigma}_{1}|}{|\bm{\Sigma}_{2}|}$ by $\Sigma_d$, a general classification rule/classifier, 
proposed by \cite{Bose/etc:2015}, is  given by  
\begin{eqnarray}
\label{gqda0}
\bm{ x}\in R_{1} & \mbox{ if }                                                    
	{\Delta_d}^2 \geq c \Sigma_d, \nonumber \\
		 \bm{x}\in R_{2} & \mbox{ otherwise},
\end{eqnarray}
for some constant  $c \geq 0$. 
Clearly, this classifier boils down  to the MMD and the QDA classifiers whenever $c$ is chosen to be  0 and 1, respectively.
In practice, the parameters in the classifier  (\ref{gqda0}) are unknown and need to be estimated from the training set.
The simplest and the most popular estimators of the  population  mean vector   and  the dispersion   matrix  
are the sample  mean  vector   and the sample  dispersion  matrix   i.e. $\widehat{\bm{\mu}}_j = \overline{\bm{x}}_j$ and 
$\widehat{\bm{\Sigma}}_j=\bm{S}_j$, which   are obtained from  the sample observations from the  $j$th population  in the training set,
for  $j=1, 2$. Accordingly, for any  new observation $\bm{x}$, the classification  rule 
 referred to as the GQDA classification rule/classifier in \cite{Bose/etc:2015}  is given by
\begin{eqnarray}
\label{gqda}
\bm{ x}\in R_{1} & \mbox{ if }                                                    
\widehat{\Delta}_d^2 \geq c \widehat{\bm{\Sigma}}_d, \nonumber \\
\bm{x}\in R_{2} & \mbox{ otherwise},
\end{eqnarray} 
where 
\begin{equation}
 \widehat{\bm{\Sigma}}_d= \log\left(\frac{\left|\bm{S}_{1}\right|}{\left|\bm{S}_{2}\right|}\right)   \mbox{ and } 
\widehat{\Delta}_d^2=(\bm{x}-\overline{\bm{x}}_2)'\bm{S}_{2}^{-1}(\bm{x}-\overline{\bm{x}}_2)
-(\bm{x}-\overline{\bm{x}}_1)'\bm{S}_{1}^{-1}(\bm{x}-\overline{\bm{x}}_1).
\nonumber
\end{equation} 

\cite{Bose/etc:2015} suggested to choose the threshold   $c$  with a view to maximize the resulting classification accuracy, 
i.e. to minimize the misclassification error. 
As noted by them, the constant $c$ may depend on $\bm{x}$ and so a suitable nonparametric approach 
needs to be adopted  to estimate an appropriate value of the constant $c$ from the training set itself. The major advantage of  GQDA is that 
a proper choice of $c$ makes the GQDA procedure adaptive to any data set, safeguarding its performance   against the possible  violation of 
the normality assumption in the classical QDA. Two methods, namely the  minimization of   the resubstitution (training set misclassification) error and  the cross-validation error, 
  have been  proposed by \cite{Bose/etc:2015}  for selecting the optimal value of $c$. 
For a ready reference, the minimization of the resubstitution error  based algorithm   is described below,  which will be used in our proposed RGQDA.\\

\noi\underline{\textit{Algorithm 1. Selection of $c$ in GQDA for two class classification.}}{\it
\begin{itemize}
	\item  Estimate $\Sigma_d$  by $ \widehat{\bm{\Sigma}}_d $ and obtain  $\widehat{\Delta}_d^2$ for each of the observations corresponding to each population in the training set. 
	\item In the training set, compute $\frac{ \widehat{\Delta}_d^2}{\widehat{\bm{\Sigma}}_d}$ for each observation  and   
	denote the ordered $\frac{ \widehat{\Delta}_d^2}{\widehat{\bm{\Sigma}}_d}$ values  for $n_j$ observations corresponding to $jth$ population   by  $r_{j(1)},\ldots, r_{j(n_j)}$,   
	$j=1, 2$. 
		 \item If $r_{2(n_2)}< r_{1(1)}$ (i.e. the sets of $\frac{ \widehat{\Delta}_d^2}{\widehat{\bm{\Sigma}}_d}$ values for the two populations in the training set 
	 are completely disjoint), do: 
		 \begin{itemize}
		 	\item Take  $c$ to be equal to any point  in the interval [$r_{2(n_2)},r_{1(n_1)}$] resulting in the 
		 	 resubstitution error to  be zero.
		 \end{itemize}
	 
	 \item  If $r_{2(n_2)}> r_{1(1)}$ (i.e. the sets of  $\frac{ \widehat{\Delta}_d^2}{\widehat{\bm{\Sigma}}_d}$ values for the two populations in the training set  overlap),
	 do:
	 \begin{itemize}
%	 	\item  then the determination of the optimal value of $c$ needs more careful scrutiny of the ${\Delta_d}^2/\Sigma_d$ values. In this case, 
		\item Find $s \geq1$, such that $r_{2(n_2-s+1)},\ldots, r_{2(n_2)}$  all exceed $r_{1(1)}$.
		\item Choose  these $s$ values as candidate values of $c$. 
		\item    Compute the resubstitution error for each of these $s$ values.
		\item Choose that value of $c$ for which the resubstitution error is the minimum and denote it by $c^*$. 
		\item If $c^*>1$, set $c^*= 1$. 
	 \end{itemize}
\end{itemize}
}

\subsection{ Multi-class classification problem}

In a general $g$-class classification problem where $g>2$,  an object  is classified into one of the $g$ populations /classes, say, $P_{1},P_2,\ldots,P_{m}$ using a decision rule  formed on the basis of an observation   $\bm{x}=(x_1,\ldots,x_p)'$ on a random feature vector $\bm{X}$.  It is assumed that  $\bm{X}$  has a probability density function   $f_i(\bm{x})$ in the class $P_i$, $i=1,2,\ldots,g$.   The feature  space $\cal X$ is essentially partitioned into $R_{1},R_2,\ldots,R_g$ such that  the object is classified into $P_j$ if $\bm{x} \in R_j$. Under the assumption of equality of the prior probabilities $\pi_1,\pi_2,\dots,\pi_m$, where $\sum_{j=1}^m \pi_j=1$,  the Bayes rule sets
	$$R_{i}=\left \{\bm{x}:\frac{f_{i}(\bm{x})}{f_{j}(\bm{x})}\geq 1\right \}=\left \{\bm{x}:\log\frac{f_{i}(\bm{x})}{f_{j}(\bm{x})}\geq 0, \ \forall j\neq i\right \},\;\;i=1,2,\ldots,g.$$
If the $g$ underlying populations are  multivariate Normal  with the mean  vector  $\bm{\mu_{i}}$ and the dispersion   matrix $\bm{\Sigma}_i$ for the $i$th population, $i=1,2,\ldots,g$, 
	 \begin{eqnarray}
\log\frac{f_{i}(\bm{x})}{f_{j}(\bm{x})}
%&=&\frac{1}{2}\!\log\left(\frac{|\bm{\Sigma}_j|}{|\bm{\Sigma}_i|}\right)+\frac{1}{2}\left [(\bm{x}-\bm{\mu_{j}})'\bm{\Sigma}_{j}^{-1}                                                  (\bm{x}-\bm{\mu_{j}}) \nonumber 
%  -(\bm{x}-\bm{\mu_{i}})'\bm{\Sigma}_{i}^{-1}(\bm{x}-\bm{\mu_{i}}) \right ] \nonumber \\
&=&\frac{1}{2}\!\log\left(\frac{\left|\bm{\Sigma}_{j}\right|}{\left|\bm{\Sigma}_{i}\right|}\right)+\frac{1}{2}\Delta_{ij}^2,
%\mbox{ say},
\nonumber
\end{eqnarray}
where
$\Delta^2_{ij}=(\bm{x}-\bm{\mu_{j}})'\bm{\Sigma}_{j}^{-1}(\bm{x}-\bm{\mu_{j}})-(\bm{x}-\bm{\mu_{i}})'\bm{\Sigma}_{i}^{-1}(\bm{x}-\bm{\mu_{i}}).$ 
This leads to the partition of the feature space as 
\begin{eqnarray}
\label{1}
R_{i} &=&\left \{\bm{x}: \frac{1}{2}\log\left(\frac{\left|\bm{\Sigma}_{j}\right|}{\left|\bm{\Sigma}_{i}\right|}\right)+\frac{1}{2}\Delta^2_{ij}\geq 0, \ \forall j\neq i \right \}\nonumber \\
& = & \left \{\bm{x}:\Delta^2_{ij} \geq \log\left(\frac{\left|\bm{\Sigma}_{i}\right|}{\left|\bm{\Sigma}_{j}\right|}\right), \ \forall j\neq i \right\}.
\end{eqnarray}

\cite{Bose/etc:2015}  proposed an extension of  the GQDA classifier (\ref{gqda}) for the general $g$-class classification problem 
which is described below.\\
 
\noi\underline{\textit{Algorithm 2. Selection of $c$ in GQDA for multi ($g >2$) class classification.}}{\it
\begin{itemize}
\item Define
$u_{ij}(\bm{x})=\frac{\Delta^2_{ij}}{\Sigma_{d_{ij}}},$
where $\Sigma_{d_{ij}}= \log\left(\frac{\left|\bm{\Sigma}_{j}\right|}{\left|\bm{\Sigma}_{i}\right|}\right)$,
 for $i, j = 1, 2, \ldots, g$.

\item Estimate $\Sigma_{d_{ij}}$  by $\widehat{\bm{\Sigma}}_{d_{ij}} $ and obtain  $\widehat{\Delta}_{d_{ij}}^2$ 
for each of the observations corresponding to each ordered pair of populations $(P_i, P_j)$,
with $i,j \in \{1,2,\ldots,g\}$ but $i\ne j$, 
by replacing the population  mean  vectors and  the dispersion   matrices by the corresponding sample counterparts  in the training set. 
	
\item Compute and subsequently order  $\widehat{u_{ij}}(\bm{x})= \frac{\widehat{\Delta}^2_{ij}}{\widehat{\Sigma}_{d_{ij}}} $ 
for each training sample $\bm{x}$ from the populations  $P_i$ and $P_j$, for each ordered pair of populations $(P_i,P_j)$,
with $i,j \in \{1,2,\ldots,g\}$ but $i\ne j$.
		
\item Let $T=\left\{\widehat{u_{ij}}(\bm{x}):\widehat{ u_{ij}}(\bm{x})\in [0,1],\ i,j \in \{1,2,\ldots,g\}, ~ i\ne j \right\}$.

\item Take each $t\in T$ as a candidate value of $c$  and correctly classify the training samples from the population $P_i$  
by comparing  $\widehat{u_{ij}}(\bm{x})$ with $c$ for each $j\ne i$. 

\item For each $j\ne i\in\{1, 2, \ldots, g\}$, define the set of correctly classified training samples from $P_i$ identified through such comparisons as
$$
R_{ij}(c)=\left\{\bm{x}\in D_{1i}:\widehat{ u_{ij}}(\bm{x})\geq c\right\}, 
%\mbox{ where } c\in T,
$$
where $c\in T$ and $D_{1i}$ is the training set for the population $P_i$.

\item Identify the number of misclassified training samples from $P_i$ as
 $$MC_i(c)=n_i-|\bigcap_{j \ne i} R_{ij}(c)|,$$
where $n_i=|D_{1i}|$ is total number of training samples from the population $P_i$.

\item Repeat the  procedure  for $i=1,2,\ldots,g$ to obtain the number of misclassified training samples from $P_i$. 

\item Compute  the total number of  instances of misclassification with a particular choice of $c$ as $ MC(c)= \sum_{i=1}^g MC_i(c).$  
\item The optimal value  of $c$ is taken as  $c^*=\text{arg}\min\limits_{c \in T}\  MC(c).$
\end{itemize}
}
%The definition of  $c_{test}$ is extended analogously following  the two-class classification problem.

\subsection{Non-robustness }
\label{SEC:nonrobust}

In reality, quite often the data get contaminated by the presence of outliers. 
It has been observed by many researchers in recent decades that, under contamination in the data, 
both the LDA and the QDA procedures result in  significantly large misclassification errors failing to provide an appropriate inference on classification, 
 mainly because of the extreme non-robustness nature of the sample mean  vector  and the sample  dispersion   matrix used in the process.
Since  the GQDA classifier also utilizes the  sample  mean  vector  and the sample dispersion   matrix, 
it is also non-robust in the presence of  outliers in the sample data and, as a result, fails to perform satisfactorily. 
This issue is illustrated below through a small simulation study pertaining to the two class problem;
more numerical illustrations are provided in the later sections of the paper.

%\bigskip
%\noindent
%\textbf{A Simulation Example:}\\
%**** Add Table 1 *****\\

Table \ref{tab1} presents the results from a simulation exercise with 10000 observations, 
of which 5000 observations are chosen from  a tri-variate Normal {\bf N}$_3$($-${\bf 1}$^\prime$, {\bf I }) to form class 1, 
and the remaining  5000  observations are chosen from  {\bf N}$_3$({\bf 1}$^\prime$, 2{\bf I }) to form   class 2. 
Here, as usual,  we denote by {\bf 1}  and  {\bf I }, the vector of all ones and  the identity matrix,  respectively. 
For each of the two classes so defined, a random selection of 1000 observations form  a training set 
and the remaining 4000 observations form  the test set, to be used to check the validity of the  classification rule. 
Under this set-up, the  value of the threshold $c^*$ of the GQDA classifier is obtained following Algorithm 1 on the pooled training  set of size 2000  
 and the \% misclassification  error  (ME\%)  is  calculated on the test data. 
For comparison, we have also worked solely with  the pooled test data set of size 8000 to obtain another threshold,  denoted by $ c_{test} $, 
which will ideally minimize the test set misclassification  error and 
help us to examine how closely it can be approximated by $c^*$, obtained on the basis of the training set only.

%\begin{table}[h!]
\begin{table}[!th]
\centering
\caption{Non-robustness of GQDA under contamination}\label{tab1}
\resizebox{0.9\textwidth}{!}{
	\begin{tabular}{|c|c|c|c|c|c|c|}
        \hline
\multicolumn{1}{|c|} {Contaminated part }& \multicolumn{2}{c|}{ Contamination } &\multicolumn{1}{c|}{$c^*$}&\multicolumn{1}{c|}{ME\% with $c^*$}&\multicolumn{1}{c|}{$c^*_{test}$ from }&\multicolumn{1}{c|}{ME\% with $c^*_{test}$} \\
\multicolumn{1}{|c|} {  of the data}& \multicolumn{2}{c|}{type and degree} & & \multicolumn{1}{c|}{on the test data}  &\multicolumn{1}{c|}{ the test data }& \multicolumn{1}{c|}{on the test data}\\
 \hline
	&\multicolumn{2}{|c|}	{Nil}    &		1 	&6.938 	&1.005 &	6.728 \\ \hline
	Train  & {\it mild } & 5\%  &		0.576 &	11.31 &	0.997 &	6.78 \\ \cline{3-7}
	& &           10\% &	0.471 &	13.189 &	0.974 &	6.793\\ \cline{3-7}
& &            15\% 	&0.413 &	13.969 &	0.979 &	6.795 \\ \cline{3-7}
 & &           20\% 	&0.375 &	13.823 &	0.981 &	6.726 \\ \hline
Train				 & {\it hard } & 5\%	&	0.437 	&25.541 	&0.996 &	6.782 \\ \cline{3-7}
				 &  & 10\%	&	0.754 &	37.826 &	1.039 &	6.759 \\ \cline{3-7}
&	& 15\% 	&0.838 &	41.254 &	0.974 &	6.694 \\ \cline{3-7}
&  & 20\%		&0.919 	&43.194 	&0.941 &	6.691 \\ \hline 
	 Train and Test &	{\it mild }& 5\%	&	0.589 &	10.849 	&0.598 	&10.444 \\ \cline{3-7}
	& &            10\% 	&	0.491 &	12.462 &	0.472 &	12.101 \\ \cline{3-7}
	&	&            15\% &	0.411 &	12.363 &	0.419 	&12.245 \\ \cline{3-7}
	& &           20\%		&0.373	&11.775	&0.373 	&11.511\\ \hline
Train and Test		 &   {\it hard } & 5\%	&	0.437	&27.061	&0.413 &	26.626\\ \cline{3-7}
		& &           10\% &	0.734&	38.007 	&0.741 &	37.514 \\ \cline{3-7}				
				& & 15\%  &	0.877	&38.785 &	0.847 	&38.287 \\ \cline{3-7}
			& & 20\% 	&0.936 &	37.775 &	0.913 &	37.456\\ \hline				
				\end{tabular}
			}
	\end{table}

Now to check the performance of the GQDA classifier in the presence of contamination, 
we serially replace  5\%, 10\%, 15\% and 20\%  observations of the training  set only, without disturbing the test set 
and also  of both the  training and the test  set separately, with observations  generated from two different tri-variate  Normal distributions 
with parameter values set widely apart form those of the original populations. 
Thus the presence of observations from the outlying populations  will contaminate the original data set, 
the contaminations being  termed as \textit{mild (hard)} depending on whether 
the  mean  vector of the outlying population is along the same (opposite) direction of that of the original population.  
The required  percentage of  {\it mild }  contamination  for the class 1 and the class 2  are   induced 
with observations generated  from  {\bf N}$_3$(-9{\bf 1}$^\prime$, 4{\bf I })  and  {\bf N}$_3$(9{\bf 1}$^\prime$, 16{\bf I }), respectively. 
Analogously, the required  percentage of {\it  {\it hard } } contamination  is induced with the observations generated 
from  {\bf N}$_3$(9{\bf 1}$^\prime$, 4{\bf I }) and {\bf N}$_3$(-9{\bf 1}$^\prime$, 16{\bf I}) for the  class 1 and the class 2,  respectively. 
We apply the GQDA classification rule  on the contaminated data sets  and 
obtain  the \% misclassification error (ME\%)  corresponding  to $c^*$ and $c_{test}$ thus emerged.   
The  above procedure is repeated  500 times   in the absence as well as in the  presence of each type and degree of contamination,
and the average \% misclassification errors are noted in Table 1.

From Table \ref{tab1}  it is seen  that,  in the absence  of contamination, the
GQDA  classification rule performs quite well as the  value of $c^*$(1.000)  and  
the \% misclassification errors  (6.938\%)  are very close to the  $c_{test} $ (1.005) 
and the corresponding \% misclassification error (6.728\%). 
Ideally, no threshold on the test  set  can perform better than $c_{test} $ as long as the test set is {\it pure}, i.e. devoid of any contamination. 
Thus, in the {\it pure} scenario the justification of the  GQDA classifiaction rule is reestablished, as has been observed  by \cite{Bose/etc:2015}.

However, in the presence of contamination, no matter whether it is  {\it mild} or  {\it hard}, 
the performance  of the GQDA classification rule starts worsening, suggested by the higher \% misclassification errors 
as the degree of contamination graduates. Whenever  only the training  set is even {\it mildly} contaminated, 
the \% misclassification errors vary from 11.31\% to  13.969\%, which is  double or more, compared to the  ones observed in the {\it pure } data set. 
In the case of {\it {\it hard }} contamination, the same is four fold or more, establishing the fact that the
GQDA classification rule is highly unsuitable in the presence of outliers. 
It is obviously expected that as long as only  the training   set is contaminated, 
it will not affect the choice of $c_{test} $ and the corresponding \% misclassification error, which is also clear from our simulation study. 	
In such  cases, irrespective of  the \textit{mild}  or \textit{hard} contamination,  $c_{test} $  is 	 1.00 approximately 
and the \% misclassification errors are close to  6.728\%, the one obtained  
when the GQDA classification rule is applied on  the {\it pure} test  set, as depicted  in the first row of  Table\ref{tab1}.
	
%\noindent 
Whenever both the train  and the test  sets are contaminated separately with the same type and degree   of contamination, 
we find that the GQDA  classification rule behaves miserably with respect to both  $c^*$ and $c_{test} $, failing  to capture the nature of the data set. 
With $c^* $,   the \% misclassification errors vary from 10.849\% to 12.462\%  (nearly doubled) 
and from 27.061\% to 38.785\% (nearly four folds)  in cases of \textit{mild}  and \textit{hard} contamination, respectively, 
suggesting  that the GQDA  classification rule  is not at all reliable for classification in the presence of outliers,  
unlike the accuracy which  transpires in the  case of {\it pure} data. 
This motivates us   to re look for a robust   classification rule   to discriminate between two elliptically symmetric distributions 
which are not necessarily Normal.

%\newpage
\section{ RGQDA: robust version of GQDA}
\label{SEC:RGQDA}

In our present work, we take care of the possibility of contamination of populations by outlying observations 
and propose to replace the classical estimators of  the  mean  vector  and  the dispersion   matrix used in  the GQDA classifier (\ref{gqda}), 
by different types of robust estimators available in the literature. 
It has been observed that the \% misclassification error  reduces drastically  using  the new version of the   classifiaction rule. 
The proposed robust procedure will be referred to as robust generalized discriminant analysis (RGQDA) hereafter,  
and a comparative study on the performances  of the GQDA classifier using  different types of  robust estimators   will be presented in the subsequent sections. 
Before that, for the sake of completeness, we present  below a brief summary of the different types of robust estimators of  the mean  vector and the dispersion   matrix 
considered by us in the  determination of the threshold $c$ in the RGQDA classifier. 

In the following, 
we  assume that the $p$-dimensional sample observations $\boldsymbol{x}_1, \ldots, \boldsymbol{x}_n$ drawn from a multivariate distribution
are independent and identically distributed with  a common  mean vector  $\boldsymbol{\mu}$
and a common  dispersion   matrix $\boldsymbol{\Sigma}$. 
Recall that the classical estimators  of $\boldsymbol{\mu}$ and  $\boldsymbol{\Sigma}$ 
are the sample  mean  vector $\overline{\boldsymbol{x}} = n^{-1} \sum_{i=1}^{n} \boldsymbol{x}_i$ 
and the sample  dispersion   matrix 
$\boldsymbol{S}=n^{-1} \sum_{i=1}^{n} \left(\boldsymbol{x}_i-\overline{\boldsymbol{x}}\right)^2$, respectively.
Both of these classical estimators  are extremely non-robust having zero breakdown point and unbounded influence function,
inspite of being  the  most efficient under a multivariate Normal model.

\bigskip
\noindent
\textbf{Winsorized (W) Estimator:}\\
The first type of the  robust estimator we have considered is the simple Winsorized estimators of $\boldsymbol{\mu}$ and  $\boldsymbol{\Sigma}$.
These were first proposed by \cite{Bickel:1965} by extending the idea of univariate winsorization
in the multivariate context and later discussed by  \cite{Zuo/Cui:2004}. To compute these estimators, a Winosorized sample is formed by 
trimming a certain percentage of observations  from the top and the bottom of the sample, 
taking into consideration of the shape of the  distribution of the data. 
But, unlike the usual trimming approach which removes the trimmed observations of the sample for subsequent analysis, 
in Winsorization the  trimmed values in the lower and the upper end of the original data are replaced respectively, by the  lowest and highest data points  of the remaining  untrimmed data.
We have used the R-function \textit{`winsor'} from package \textit{`psych'} to form the Winsorized sample
and then use the usual  mean  vector  and   the dispersion   matrix of the Winsorized sample as  robust Winsorized estimators of 
$\boldsymbol{\mu}$, and $\boldsymbol{\Sigma}$, respectively.

%\bigskip
%\noindent
%\textbf{Winsorized Modified One-Step M (WMOM) estimator:}\\
%A set of modification of the simple Winsorized estimators of $\boldsymbol{\mu}$ and  $\boldsymbol{\Sigma}$ 
%was developed by \cite{Wilcox/Keselman:2003} using a modified one-step M-estimators.
%Here,  the Winsorized sample (say $\underline{\boldsymbol{X}}_{W}$) is prepared as before, and then the WMOM estimators of 
% $\boldsymbol{\mu}$ and  $\boldsymbol{\Sigma}$ are defined respectively as 
%the the mean vector vector  of $\underline{\boldsymbol{X}}_{W}$ and the product of Spearman's correlation coefficient and $\boldsymbol{M}_n \boldsymbol{M}_n ^T$, 
%where $\boldsymbol{M}_n$ is the component-wise (scaled) median absolute deviation (MAD) of $\underline{\boldsymbol{X}}_{W}$; 
%see \cite{Haddad/etc:2013} for mathematical details.
%These modified estimators have been successfully applied to several real life problems and 
%are found to be computationally as fast as our standard (non-robust) estimation process unlike the sophisticated robust estimators described below. However the properties of these estimators are not yet well established theoretically. 

\bigskip
\noindent
\textbf{Minimum Volume Ellipsoid (MVE) Estimator:}\\
Robust estimators of $\boldsymbol{\mu}$ and  $\boldsymbol{\Sigma}$ having high breakdown property, called
the minimum volume ellipsoid (MVE) estimators, were proposed by \cite{Rousseeuw:1985}.
These estimators are  defined  on the basis of a  sample  of  fixed size $h (<n)$, 
which lies within an ellipsoid of minimum volume, taking the center and the spear of that ellipsoid as the MVE estimators of $\boldsymbol{\mu}$
and  $\boldsymbol{\Sigma}$, respectively. For a suitable choice of $h$, 
one can achieve  a very high breakdown point of the MVE estimators \citep{Davies:1987},
which makes these estimators widely popular besides their simple interpretation.
However, the MVE estimators are not $\sqrt{n}$-consistent \citep{Davies:1992a} and hence not efficient.
We have used the R function \textit{`CovMve'} from package \textit{`rrcov'} for computation of the MVE estimators.

\bigskip
\noindent
\textbf{Minimum Covariance Determinant (MCD) Estimator:}\\
\cite{Rousseeuw:1985} also proposed another type of robust estimators of $\boldsymbol{\mu}$
and  $\boldsymbol{\Sigma}$ by the sample  mean  vector  and the dispersion   matrix of $h(<n)$ sample observations 
which  leads to the minimum value of the determinant of the  dispersion  matrix   over all such samples of fixed size $h$.
Hence these estimators are known as the minimum covariance determinant (MCD) estimators.
The MCD estimators with $h=[(n+p+1)/2]$ achieve the highest possible (finite-sample) breakdown point
among the class of affine equivariant scatter estimators \citep{Davies:1987}.
Further, MCD estimators have bounded influence functions as described in \cite{Croux/Haesbroec:1999},
but they do not have very high efficiency. 
Unlike the MVE estimators, the MCD estimator of $\boldsymbol{\mu}$ has been  shown to be $\sqrt{n}$-consistent 
and asymptotically Normal by \cite{Butler/etc:1993}.
There are several fast algorithms available for computation of this popular estimator.
We have used the R function \textit{`covMcd'} in our implementation 
that utilizes a fast MCD computation algorithm proposed by \cite{Rousseeuw/vanDriessen:1999}.

\bigskip
\noindent
\textbf{M-Estimator:}\\
The M-estimators of  $\boldsymbol{\mu}$ and  $\boldsymbol{\Sigma}$
are defined by \cite{Maronna:1976} as the respective solutions $\boldsymbol{l}_n\in \mathbb{R}^p$ and $\boldsymbol{V}_n$,
a positive definite matrix, of the system of estimating equations
\begin{eqnarray}
\frac{1}{n} \sum_{i=1}^n 
\psi_1\left(\sqrt{(\boldsymbol{x}_i - \boldsymbol{l}_n)^T\boldsymbol{V}_n^{-1}(\boldsymbol{x}_i - \boldsymbol{l}_n)}\right)
(\boldsymbol{x}_i - \boldsymbol{l}_n)
&=& \boldsymbol{0}_p,
\nonumber\\
\frac{1}{n} \sum_{i=1}^n 
\psi_2\left({(\boldsymbol{x}_i - \boldsymbol{l}_n)^T\boldsymbol{V}_n^{-1}(\boldsymbol{x}_i - \boldsymbol{l}_n)}\right)
(\boldsymbol{x}_i - \boldsymbol{l}_n)(\boldsymbol{x}_i - \boldsymbol{l}_n)^T
&=& \boldsymbol{V}_n,
\nonumber
\end{eqnarray}
for two suitably given weight functions $\psi_1$ and $\psi_2$.
%Besides proposing these estimators, 
\cite{Maronna:1976} also derived their detailed asymptotic (consistency and normality) and robustness (influence function and breakdown point) properties.
In particular, with suitable choice of $\psi_i$s, these M-estimators are highly efficient under Normal model
and locally robust having bounded influence function, but they do not have high breakdown point in higher dimensions.
Note that the maximum likelihood estimators of $\boldsymbol{\mu}$ and  $\boldsymbol{\Sigma}$ 
under any radically symmetric location-scale model density are also  M-estimators 
for   proper choices of weight functions $\psi_i$s.
The most commonly suggested weight functions for robust inference are those proposed by Huber \citep{Huber:1981} as given below
$$
\psi_1(z) =\max(-k, \min(z,k)), ~~~
\psi_2(z) = \frac{\max(-k^2, \min(z,k^2))}{E_{\boldsymbol{X}\sim N_p(\boldsymbol{0}_p, \boldsymbol{I}_p)}\left[\max(-k^2, \min(||\boldsymbol{X}||^2,k^2))\right]},
$$
for a given tuning parameter $k$.  
See \cite{Hampel/etc:1986} for further details.
In this paper, we have used the R function `mvhuberM' from package `SpatialNP'
for the computation of these M-estimators with the above-mentioned Huber's weight functions.

\bigskip
\noindent
\textbf{S-Estimator:}\\
S-estimators of $\boldsymbol{\mu}$ and  $\boldsymbol{\Sigma}$  are  smoother extensions of the MVE estimators
so that the resulting estimators become $\sqrt{n}$-consistent and asymptotically Normal.
For a given non-negative, symmetric and continuously differentiable function $\rho$ with $\rho(0)=0$
and a constant $c< \sup \rho$, the corresponding S-estimators of $\boldsymbol{\mu}$ and $\boldsymbol{\Sigma}$
are defined by \cite{Davies:1987} as the respective solutions $\boldsymbol{l}_n\in \mathbb{R}^p$ and $\boldsymbol{V}_n$,
a positive definite matrix, of the constrained optimization problem 
\begin{eqnarray}
\label{opt}
\min \left|\boldsymbol{V}_n\right|,  
~~~\mbox{subject to }~
\frac{1}{n} \sum_{i=1}^n 
\rho\left(\sqrt{(\boldsymbol{x}_i - \boldsymbol{l}_n)^T\boldsymbol{V}_n^{-1}(\boldsymbol{x}_i - \boldsymbol{l}_n)}\right)
\leq c.
%\nonumber
\end{eqnarray}
%discussed in \cite{Davies:1987}.
Further, to achieve robustness one needs to ensure that
there exists a constant $b>0$ such that the function $\rho$ is strictly increasing in $[0, b)$, being 
constant on $[b, \infty)$. 
Depending on the choice of the constant $c$ in the above optimazation problem (\ref{opt}), the S-estimators can have  either high efficiency under a Normal model
or the maximum possible (finite sample) breakdown point among all affine equivariant estimators,
but not the both simultaneously.
But S-estimators always have bounded influence function indicating their local robustness
and are also closely related to the M-estimators as explored by \cite{Lopuhaa:1989}.
A popular choice of the function $\rho$  is  the biweight function proposed by Tukey  defined as 
$$
\rho^{\mbox{Tukey}}(s) = s\left(1-\frac{s^2}{b^2}\right)^2 I(s<b), ~~\mbox{for } s>0. 
$$
A fast deterministic algorithm for the computation of S-estimators,   proposed in \cite{Hubert/etc:2012},
is used in our present work through the R-function \textit{`CovSest'} from package \textit{`rrcov'}.
%For all our illustrative examples with S-estimators, the required breakdown point is kept fixed at $0.75$.
%CovSest(xtest1,bdp=0.75,method="sdet"); the required breakdown point. bdp=0.75;
%The method 'sdet' invokes the deterministic algorihm of Hubert et al. (2012).\\
%Computes S-Estimates of multivariate location and scatter based on Tukey's biweight function using a fast algorithm similar to the one proposed by Salibian-Barrera and Yohai (2006) for the case of regression. Alternativley, the Ruppert's SURREAL algorithm, bisquare or Rocke type estimation can be used.

\bigskip
\noindent
\textbf{Stahel-Donoho (SD) Estimator:}\\
\cite{Stahel:1981} and \cite{Donoho:1982} proposed other interesting robust estimators of $\boldsymbol{\mu}$ and $\boldsymbol{\Sigma}$, 
which are affine equivariant as well as have high breakdown point in higher dimensions.
They defined a multivariate ``outlyingness" measure of a point $\boldsymbol{x}\in \mathbb{R}^p$
with respect to the observed sample (say, $\underline{\boldsymbol{X}}_n=\{\boldsymbol{x}_1, \ldots, \boldsymbol{x}_n\}$)
as 
\begin{eqnarray}
O(\boldsymbol{x}, \underline{\boldsymbol{X}}_n)
= \sup\limits_{\{\boldsymbol{u}\in \mathbb{R}^p : ||\boldsymbol{u}||=1 \}}
\frac{\left|\boldsymbol{u}'\boldsymbol{x} - \mu_n(\boldsymbol{u}\cdot \underline{\boldsymbol{X}}_n)\right|}{
\sigma_n(\boldsymbol{u}\cdot \underline{\boldsymbol{X}}_n)},
\nonumber
\end{eqnarray}
where $\mu_n(\boldsymbol{u}\cdot \underline{\boldsymbol{X}}_n)$ and $\sigma_n(\boldsymbol{u}\cdot \underline{\boldsymbol{X}}_n)$
are some given estimators of the univariate  mean  and the standard deviation of the sample 
$\boldsymbol{u}\cdot \underline{\boldsymbol{X}}_n 
=\left\{\boldsymbol{u}'\boldsymbol{x}_1, \ldots, \boldsymbol{u}'\boldsymbol{x}_n\right\}$ for any $\boldsymbol{u}\in \mathbb{R}^p$.
Common examples of the univariate estimator of  $\mu_n (\sigma_n)$ are the  mean vector or the Median or any general M-estimator of the univariate location (standard deviation or the mean  absolute deviation 
(MAD)  or  any general M-estimator of the univariate scale parameter) which can be used in $O(\boldsymbol{x}, \underline{\boldsymbol{X}}_n)$; 
see \cite{Hampel/etc:1986}.
The Stahel-Donoho (SD) estimators of  $\boldsymbol{\mu}$  and $\boldsymbol{\Sigma}$
based on the sample data $\underline{\boldsymbol{X}}_n$ are then defined respectively, 
as the  weighted  mean   vector and  the dispersion  matrix,  given by
\begin{eqnarray}
\boldsymbol{l}_n^{SD} = 
\frac{\sum\limits_{i=1}^{n} w\left(O(\boldsymbol{x}_i, \underline{\boldsymbol{X}}_n)\right)\boldsymbol{x}_i}{
	\sum\limits_{i=1}^{n} w\left(O(\boldsymbol{x}_i, \underline{\boldsymbol{X}}_n)\right)},
~~
\boldsymbol{V}_n^{SD} = \frac{\sum\limits_{i=1}^{n} w\left(O(\boldsymbol{x}_i, \underline{\boldsymbol{X}}_n)\right)
	\left(\boldsymbol{x}_i - \boldsymbol{l}_n^{SD}\right)\left(\boldsymbol{x}_i - \boldsymbol{l}_n^{SD}\right)^T}{
	\sum\limits_{i=1}^{n} w\left(O(\boldsymbol{x}_i, \underline{\boldsymbol{X}}_n)\right)},
\nonumber
\end{eqnarray}
where $w$ is a suitable weight function resulting in outlier downweighting.
By choosing affine equivariant and high breakdown univariate estimators  of $\mu_n$ and $\sigma_n$,
one can simultaneously achieve the affine equivariance and  high breakdown point for the resulting 
SD estimators of $\boldsymbol{\mu}$ and $\boldsymbol{\Sigma}$ as well. 
The $\sqrt{n}$-consistency of these SD estimators was proved by \cite{Maronna/Yohai:1995},
but their asymptotic distributions are   yet to be determined.
We have used the R-function \textit{`CovSde'} from the package \textit{`rrcov'} 
for the computation of the SD estimators with the extreme 5\% observations getting zero weights 
through their outlyingness  measure, computed using   the univariate median and the MAD combination.

We have briefly pointed out  that  all the above robust estimators of $\boldsymbol{\mu}$ and $ \boldsymbol{\Sigma}$
have some advantages and disadvantages  and  none is the uniformly best. 
In the next section, we will empirically compare the performances of the RGQDA classifiers  formed by replacing 
the   classical estimators of the   mean vector and the  dispersion  matrix  in   the GQDA classifier by the six above-mentioned robust estimators.

\section{Empirical illustrations: simulation studies}
\label{SEC:simulation}

\subsection{Experimental set-ups }

In order to investigate the effectiveness of the proposed RGQDA over GQDA, we have conducted several simulation studies.
Due to the equivariance of the estimators used, the results obtained are mostly similar and hence, for brevity, 
we only present the results for one two-class  problem and another multi-class problem with four classes,
each for three different elliptical symmetric distributions, as specified below.

\begin{itemize}
	\item\textbf{Two-class Problem:} We consider the simulation set-up as described in Section \ref{SEC:nonrobust} with Normal distribution. 
	As mentioned there, 500 repetitions of the simulation study have been undertaken for each case under the {\it pure} data as well as under the different types 
	and magnitudes of data contaminations.  Here, we also carry out the similar simulation studies for 
	the $t$-distribution with 3 degrees of freedom (d.f.) and the heavy-tailed Cauchy distribution, 
	inducing {\it mild } and {\it hard } contamination from the outlying population, having the same  mean  (location) vector and the  dispersion   (scale) matrix 
	as the case of simulation with Normal distribution.

	\item\textbf{Four-class Problem:} As before, we consider three types of elliptically symmetric distributions, 
	namely multivariate (6-variate) Normal, $t$ with 3 degrees of freedom and Cauchy distribution, with the same size  of the  train and the test sets. 
	Observations in the four classes are generated from the four postulated distributions having the same  dispersion  matrix $\Sigma=\bf{I}_6$
	but with different  mean  vectors, given by  $\boldsymbol{\mu}_1=(1,1,1,1,1,1)'$, 	
	$\boldsymbol{\mu}_2=(1,1,1,-1,-1,-1)'$, $\boldsymbol{\mu}_3=(-1,-1,-1,-1,-1,-1)'$ and $\boldsymbol{\mu}_4=(-1,-1,-1,1,1,1)'$, respectively.
	The  outliers in the $i$-th class for $i=1, \ldots, 4$, are generated from the same distribution as the postulated one for the $i$-th class,
	but with the different  mean vectors and the dispersion  matrices; the mean vectors being $\boldsymbol{\mu}^0_i=9\times\boldsymbol{\mu}_i$ 
	and $\boldsymbol{\mu}^0_i=-9\times\boldsymbol{\mu}_i$ for the \textit{{\it mild }} and \textit{{\it hard }} contaminations, respectively,
	but the  dispersion  matrices for both  the types of contaminations  are taken as $\Sigma^0=4\Sigma$. 
	Both the types of contaminations are considered for 5\%, 10\%, 15\% and 20\% observations 
	from the train set only,  without disturbing the test set, as well as from both the train set  and the test set separately, as in the case of two-class problem.
\end{itemize}

In each of the cases, we report the average \% misclassification  error   and the corresponding standard deviation (SD) over 500 replications
obtained from the test set using the GQDA and different RGQDA  classifier cut-offs, obtained from the respective train set. 
%For brevity, the average values of $c^*$ obtained in each case are presented in the Supplementary material. (****this  supplementary material has to be included in our paper) 

\subsection{Performances under {\it pure}  data}

The results for the uncontaminated cases i.e. {\it pure} data sets are reported in Table \ref{tab Pure} for both  the two-class and four-class problems 
and the three types of distributions. One can clearly observe that for the light tailed Normal distribution, 
the performance of the GQDA classifier is the best but its different robust versions proposed in RGQDA also produce comparable  average ME\% under the {\it pure} data.
This is quite natural to expect since the sample estimates of the mean vector and the dispersion  matrix are most efficient under  a properly specified model with no contamination.
However, for the heavy-tailed $t$ and Cauchy distributions, even under the {\it  pure} data, 
several  robust  versions of the  GQDA  classifier proposed in RGQDA outperform the  GQDA classifier, with greater improvements for the Cauchy distribution,
which  is due to the presence of potential extreme observations in  these heavy tailed distributions.
Among the different robust versions, the one with either  MCD or SD works the  best for $t$- and  Cauchy distribution.

\begin{table}[!h]
\centering
\caption{ Average ME\% ( SD) for the  test data set using the  GQDA classifier and its different robust versions proposed in RGQDA  for the {\it pure} data}
\label{tab Pure}
\scalebox{0.7}{\begin{tabular}{|l|c|c|c|c|c|c|}
\hline
\multicolumn{1}{|c|}{  Method of} &\multicolumn{3}{c|}{Two-class problem}&\multicolumn{3}{c|}{Four-class problem}\\ 
%\cline{2-4}
classification &Normal & t with 3 d.f & Cauchy &Normal & t with 3 d.f & Cauchy\\
\hline
GQDA	& {\bf	6.938 (0.312)}	& {\bf	12.519 (0.636)} & {\bf 27.258 (5.131)} & {\bf 8.653 (0.943)} & {\bf 18.713 (1.297)} & {\bf 45.190 (4.359)}	\\
W		&	6.986 (0.298)	&	12.237 (0.417) & 20.155 (0.589) & 8.887 (0.997) & 16.997 (1.453) & 31.637 (1.596)	\\
%WMOM 		&	6.992(0.302)&	12.207(0.4)&	20.187(0.622)\\
MVE			&	6.976 (0.316)	&	12.262 (0.424) & 20.216 (0.642)	& 8.903 (0.857) & 17.143 (1.450) & 30.257 (1.727)	\\
MCD			&  	6.999 (0.317)	&	12.292 (0.448) & {\bf 20.124 (0.651)}	& 8.397 (0.790) & {\bf 16.113 (1.350)} & 28.570 (1.808)\\
M			& 	7.070 (0.328)	&	12.391 (0.478) & 20.408 (0.850)	& 7.313 (0.924) & 17.353 (1.309) & 28.677 (1.791)\\
S			&	7.077 (0.318)	&	12.304 (0.443) & 20.239 (0.588)	& 10.263 (1.047) & 17.543 (1.347) & 29.570 (1.920)\\
SD			&	7.004 (0.308)	&	{\bf 12.230 (0.430)} & 20.226 (0.644)	& 8.353 (1.063)  & 17.893 (1.539) & {\bf 27.860 (1.622)}\\
\hline
\end{tabular}}
\end{table}

\subsection{Performances under contamination: two-class problem}

The empirical values of the average ME\% and their SDs for simulated experiments with different contaminations for the two-class problems
are reported in Table \ref{tab N} through Table \ref{tab CA} for Normal, $t$ and Cauchy distribution, respectively. 
All these simulation studies emphatically establish that for all the elliptically symmetric distributions considered here, 
in the case of contaminated data, the use of the robust estimators  drastically reduces the \% of misclassification error.
In general, the improvement increases as the degree of contamination increases from 5\% to 20\%. 
As  expected, the use of robust estimators  exhibits more improvement in terms of reducing the \% misclassification error 
when there is \textit{hard} contamination compared to the  \textit{mild}  contamination.

%\begin{table}[h!]

\begin{table}[!h]
	\centering
	\caption{Average ME\% (SD) for two-class problems with contaminated Normal distribution, 
		using the GQDA classifier   and its different versions proposed in  RGQDA}\label{tab N}
	\scalebox{0.8}{\begin{tabular}{|c|c|c|c|c|c|}
			\hline
			\multicolumn{1}{|c|} {Contamination  }& \multicolumn{1}{c|}{  Method of} &\multicolumn{4}{c|}{Contamination type}\\ \cline{3-6}
			\multicolumn{1}{|c|} {  degree}& classification &  \multicolumn{1}{c|}{Train {\it {\it mild }}}  &\multicolumn{1}{c|}{ Train \it{hard}}& \multicolumn{1}{c|}{Train and Test {\it {\it mild }}} &  \multicolumn{1}{c|}{Train and Test {\it {\it hard }}}\\
			\hline
			{\bf 5\%}& {\bf GQDA} & {\bf 11.31(0.754)}	& {\bf 25.541(1.336)}	& {\bf 10.849(0.632)}	&{\bf 27.061(1.269)}\\ \cline{2-6}
			&W & 7.088(0.399)	&7.13(0.347)	&9.059(0.389)&	9.371(0.35)\\\cline{2-6}
%			& WMOM& 7.042(0.359)	& 7.105(0.304)	& 9.108(0.355)	& 9.319(0.316)\\\cline{2-6}
			& MVE &{\bf 6.994(0.323)}&	7.02(0.343)&	9.118(0.288)&	9.312(0.349)\\\cline{2-6}
			& MCD &7.01(0.314)&{\bf	6.947(0.344)}&	9.08(0.287)&	9.238(0.336)\\\cline{2-6}
			&M & 7.109(0.398)&	7.248(0.286)&	8.984(0.474)&	9.459(0.23)\\ \cline{2-6}
			& S &7.148(0.319)&	7.124(0.337)&	{\bf 9.03(0.482)}	&9.408(0.395)\\\cline{2-6}
			& SD & 7.027(0.317)&	7.019(0.338)&	9.085(0.398)&{\bf	9.306(0.325)}\\\cline{2-6}
			\hline \hline
			{\bf 10\%}& {\bf GQDA} &{\bf 13.189(0.769)}	&{\bf 37.826(0.705)}	&{\bf 12.462(0.841)}	&{\bf 38.007(0.623)}\\ \cline{2-6}
			& W &7.135(0.346)	&7.794(0.401)	&10.473(0.925)	&11.983(0.376)\\ \cline{2-6}
%			& WMOM & 7.111(0.288)	&7.716(0.376) &	10.584(0.947)	&12.036(0.351)\\ \cline{2-6}
			& MVE &7.005(0.339)	&7.072(0.344)	&11.255(0.354)	&11.503(0.388)\\ \cline{2-6}
			& MCD &{\bf 6.942(0.341)}	&7.079(0.311)	&11.207(0.341)	&11.489(0.303)\\ \cline{2-6}
			& M&7.561(0.413)	&10.088(0.582)&	{\bf 7.415(0.643)}	&{\bf 14.317(0.597)}\\ \cline{2-6}
			& S &7.048(0.299)	&7.037(0.381)	&11.126(0.648)	&11.851(0.681)\\ \cline{2-6}
			& SD & 6.998(0.336)	&{\bf 6.966(0.319)}	&11.315(0.352)	&11.412(0.286)\\ \cline{2-6}
			\hline \hline
			{\bf 15\%} & {\bf GQDA} & {\bf 13.969(0.739)}	&{\bf 41.254(0.463)}	&{\bf 12.363(0.655)}	&{\bf 38.785(0.526)}\\ \cline{2-6}
			& W &7.48(0.392)	&10.299(0.485)	&9.088(1.747)	&16.273(0.613)\\ \cline{2-6}
%			& WMOM&7.559(0.453)&	10.575(0.775)	&10.213(2.002)	&16.447(0.582)\\ \cline{2-6}
			& MVE &7.02(0.304)	&{\bf 6.987(0.355)}	&13.385(0.365)	&13.755(0.368)\\ \cline{2-6}
			& MCD &7.007(0.298)	&7.01(0.313)	&13.287(0.794)	&13.738(0.343)\\ \cline{2-6}
			& M&9.062(0.493)	&27.219(1.846)&	{\bf 7.7(0.425)}	&31.183(1.801)\\ \cline{2-6}
			& S&7.066(0.34)	&7.084(0.314)	&13.099(1.18)	&14.209(0.999)\\ \cline{2-6}
			& SD &{\bf 6.982(0.285)}	&7.043(0.348)	&13.375(0.353)	&{\bf 13.621(0.298)}\\ \cline{2-6}
			\hline \hline
			{\bf 20\%} & {\bf GQDA} &{\bf 13.823(0.733)} &	{\bf 43.194(0.627)} & {\bf 	11.775(0.692)} & {\bf	37.775(0.411)} \\ \cline{2-6}
			& W &8.843(0.573)	&31.401(1.676)&	{\bf 7.135(0.437)}	&35.175(1.457)\\ \cline{2-6}
%			& WMOM & 12.15(1.144)	&33.612(1.71)	&11.325(1.187)	&37.28(0.984) \\ \cline{2-6}
			& MVE &{\bf 6.971(0.271)}&	7.021(0.354)&	15.534(0.334)	&{\bf 15.926(0.33)}\\ \cline{2-6}
			& MCD&6.966(0.433)&	{\bf 6.973(0.289)}&	15.516(0.363)	&15.984(0.311)\\ \cline{2-6}
			& M& 10.634(0.526)&	41.337(0.674)	&8.552(0.418)	&37.505(0.484)\\ \cline{2-6}
			& S&7.064(0.318)&	7.076(0.367)&	15.344(0.958)&	16.692(1.303)\\ \cline{2-6}
			& SD& 6.999(0.295)&	7.051(0.303)&	15.403(0.339)	&15.934(0.37)\\ \cline{2-6}
			\hline
	\end{tabular}}
\end{table}

%\end{document}*************************
\begin{table}[!h]
	\centering
	\caption{Average ME\% (SD) for two-class problems with  contaminated $t$ distributions with 3 d.f, 
		using the GQDA classifier  and its different versions proposed in RGQDA  }\label{tab t3}
	\scalebox{0.8}{\begin{tabular}{|c|c|c|c|c|c|}
			\hline
			\multicolumn{1}{|c|} {Contamination  }& \multicolumn{1}{c|}{  Method of} &\multicolumn{4}{c|}{Contamination type}\\ \cline{3-6}
			\multicolumn{1}{|c|} {  degree}& classification &  \multicolumn{1}{c|}{Train {\it {\it mild }}}  &\multicolumn{1}{c|}{ Train \it{hard}}& \multicolumn{1}{c|}{Train and Test {\it {\it mild }}} &  \multicolumn{1}{c|}{Train and Test {\it {\it hard }}}\\
			\hline
			{\bf 5\%} & {\bf GQDA} &{\bf 14.411(0.962)}&	{\bf 22.503(2.528)}	&{\bf 13.728(1.296)}	&{\bf 24.426(1.76)}\\ \cline{2-6}
			& W &12.285(0.406)&	12.825(0.433)	&14.039(0.569)&	14.835(0.487)\\ \cline{2-6}
%			& WMOM& 12.284(0.459)	&12.889(0.46)&	{\bf 13.876(0.601)}	&14.837(0.42)\\ \cline{2-6}
			& MVE &12.335(0.461)&	12.365(0.492)&	14(0.494)&	14.281(0.393)\\ \cline{2-6}
			&MCD & 12.316(0.403)	&12.202(0.342)&	14.071(0.402)&	{\bf 14.255(0.503)}\\ \cline{2-6}
			& M &12.678(0.398)&	13.276(0.473)&	{\bf 13.993(0.671)}	&15.127(0.552)\\ \cline{2-6}
			& S &12.271(0.389)&	{\bf 12.304(0.432)}&	14.027(0.447)&	14.266(0.398)\\ \cline{2-6}
			&SD &{\bf 12.205(0.426)}&	12.261(0.457)	&14.056(0.466)	&14.267(0.334)\\ \cline{2-6}
			\hline \hline
			{\bf 10\%} &{\bf GQDA} & {\bf 15.098(1.277)} & {\bf 40.687(1.145)} & {\bf 14.099(1.083)} & {\bf 41.666(1.095)} \\\cline{2-6} 
			& W &12.546(0.487)	&14.658(0.768)	&14.629(1.552)	&18.153(0.6) \\\cline{2-6}
%			& WMOM &12.604(0.501)	&14.68(0.688)&	14.883(1.494)	&18.226(0.632) \\\cline{2-6}
			& MVE &12.262(0.393)&	12.304(0.482)	&15.972(0.461)&	16.312(0.415) \\\cline{2-6}
			& MCD &{\bf 12.201(0.403)}&	{\bf 12.171(0.394)}	&16.025(0.62)	&{\bf 16.207(0.392)}\\\cline{2-6}
			& M & 13.171(0.541)	&16.865(0.903)&	{\bf 13.122(1.082)}&	20.415(0.616)\\\cline{2-6}
			&S & 12.32(0.463)&	12.185(0.44)&	15.725(0.834)	&16.332(0.498) \\\cline{2-6}
			& SD &12.296(0.436)&	12.243(0.453)&	15.829(0.722)&	16.271(0.416)\\\cline{2-6}
			\hline \hline
			{\bf 15\%} & {\bf GQDA} & {\bf 15.181(1.049)} & {\bf	43.491(3.959)} & {\bf	13.628(1.158)} & {\bf	6.982(0.285)} \\\cline{2-6}
			& W &12.932(0.568)&	19.145(1.175)&	{\bf 12.386(1.021)}&	23.474(1.015)\\\cline{2-6}
%			& WMOM& 13.338(0.567)	&20.39(1.203)	&13.479(1.209)&	24.369(0.992)\\\cline{2-6}
			& MVE &12.249(0.423)&	12.365(0.407)&	17.68(0.693)	&18.401(0.494)\\\cline{2-6}
			& MCD &{\bf 12.202(0.449)}&	{\bf 12.235(0.436)}&	17.464(0.771)&	{\bf 18.315(0.44)}\\\cline{2-6}
			& M &14.341(0.634)&	31.945(2.213)&	12.758(0.592)	&35.55(1.799)\\\cline{2-6}
			& S &12.261(0.37)&	12.239(0.444)&	17.264(1.413)&	18.718(0.89)\\\cline{2-6}
			& SD &12.339(0.393)&	12.401(0.442)&	17.427(1.344)	&18.382(0.546)\\\cline{2-6}
			\hline \hline
			{\bf 20\%} & {\bf GQDA} & {\bf 14.968(1.053)} & {\bf 45.021(5.266)}&	{\bf 12.893(1.199)} & {\bf 39.898(0.703)} \\ \cline{2-6}
			& W &13.864(0.657)&	41.288(1.241)	&{\bf 11.639(0.671)}&	40.69(0.893)\\ \cline{2-6}
%			& WMOM& 16.891(0.938)&	42.426(2.229) &	15.69(1.057)&	43.032(0.584) \\ \cline{2-6}
			& MVE &{\bf 12.28(0.442)}&	{\bf 12.265(0.439)}	&19.185(1.63)&	20.397(0.794)\\ \cline{2-6}
			& MCD& 12.292(0.424)&	12.352(0.467)	&19.498(0.53)&	{\bf 20.386(0.514)}\\ \cline{2-6}
			& M &15.381(0.729)	&44.599(0.651)&	13.04(0.727)	&40.351(0.618)\\ \cline{2-6}
			& S &12.433(0.448)&	12.325(0.424)	&19.263(1.211)	&20.389(0.566)\\ \cline{2-6}
			& SD &12.434(0.427)	&12.512(0.393)&	19.307(1.416)	&20.44(0.563)\\ \cline{2-6}
			\hline
	\end{tabular}}
\end{table}

	%\end{document}
	\begin{table}[!h]
\centering
\caption{Average ME\% (SD)  for two-class problems with  contaminated Cauchy distributions,
	 using the GQDA classifier and its different versions proposed in RGQDA   }\label{tab CA}
\scalebox{0.8}{\begin{tabular}{|c|c|c|c|c|c|}
        \hline
\multicolumn{1}{|c|} {Contamination  }& \multicolumn{1}{c|}{  Method of} &\multicolumn{4}{c|}{Contamination type}\\ \cline{3-6}
\multicolumn{1}{|c|} {  degree}& classification &  \multicolumn{1}{c|}{Train {\it mild }}  &\multicolumn{1}{c|}{ Train {\it hard } }& \multicolumn{1}{c|}{Train and Test {\it mild }} &  \multicolumn{1}{c|}{Train and Test {\it hard }}\\
 \hline 
{\bf 5\%} & {\bf GQDA} & {\bf 24.868(4.423)} & {\bf	21.582(0.756)} & {\bf	25.329(4.026)} &{\bf	36.128(6.321)} \\ \cline{2-6}
& W &{\bf 19.951(0.66)}	&{\bf 21.201(0.73)}	&{\bf 20.934(0.987)}&	22.848(0.747)\\ \cline{2-6}
%& WMOM &20.181(0.623)&	21.476(0.761)&	21.062(1.001)&	22.902(0.671)\\ \cline{2-6}
& MVE &20.259(0.753)&	20.205(0.681)	&21.62(0.804)&	21.79(0.603)\\ \cline{2-6}
& MCD &20.217(0.68)&	20.252(0.526)&	21.645(0.835)&	21.831(0.477)\\ \cline{2-6}
& M&20.284(0.687)	&21.548(0.826)&	21.296(1.232)&	23.078(0.695)\\ \cline{2-6}
& S&20.096(0.548)	&20.097(0.564)&	21.47(0.738)&	21.859(0.571)\\ \cline{2-6}
& SD &20.329(0.615)&	20.302(0.612)&	21.659(0.707)&{\bf 	21.77(0.601)}\\ \cline{2-6}
\hline \hline
{\bf 10\%} & {\bf GQDA} & {\bf 23.873(3.822)} & {\bf	36.462(6.616)} & {\bf	24.81(3.156)} &{\bf	39.747(5.038)} \\ \cline{2-6}
& W &{\bf  19.939(0.677)}&	24.069(1.523)&	{\bf 20.304(1.54)}&	26.462(1.093)\\ \cline{2-6}
%& WMOM& 20.193(0.693)&	24.553(1.121)	&20.73(1.349)	&27.035(1.074)\\ \cline{2-6}
& MVE &20.244(0.591)&	20.229(0.684)	&22.803(0.958)&	{\bf 23.382(0.583)}\\ \cline{2-6}
& MCD& 20.174(0.555)	&{\bf 20.177(0.596)}	&22.839(0.927)	&23.364(0.47)\\ \cline{2-6}
&M & 20.38(0.87)	&25.596(1.507)&	20.37(1.231)	&28.032(1.062)\\ \cline{2-6}
& S &20.287(0.576)&	20.319(0.602)&	22.863(0.775)	&23.565(0.571)\\ \cline{2-6}
& SD &20.316(0.677)	&20.267(0.785)	&22.922(1.241)&	23.727(0.541)\\ \cline{2-6}
\hline \hline
{\bf 15\%} & {\bf GQDA} & {\bf 23.523(3.746)} & {\bf	36.509(6.699)} & {\bf	24.265(4.166)} &{\bf	38.447(5.004)} \\ \cline{2-6}
& W &{\bf 19.774(0.669)}&	36.029(4.374)	&{\bf 18.576(0.817)}&	37.523(3.091)\\ \cline{2-6}
%& WMOM &20.791(0.73)&	37.402(3.436)&	20.33(1.25)&	39.271(3.23) \\ \cline{2-6}
& MVE &20.414(0.766)&	{\bf 20.14(0.607)}&	24.131(1.648)	&25.154(0.825)\\ \cline{2-6}
&MCD&20.121(0.658)&	20.254(0.624)&	24.492(0.984)&	{\bf 24.989(0.601)}\\ \cline{2-6}
& M &20.572(0.607)&	43.169(2.335)&	19.339(0.918)	&43.261(2.04)\\ \cline{2-6}
& S &20.09(0.69)&	20.148(0.474)	&24.339(0.815)	&25.155(0.552)\\ \cline{2-6}
& SD&20.363(0.754)&	20.427(0.685)&	23.805(2.029)	&25.311(0.724)\\ \cline{2-6}
\hline \hline
{\bf 20\%} & {\bf GQDA} &{\bf 22.867(3.529)} & {\bf	30.835(6.096)} & {\bf	24.544(4.502)} &{\bf	35.449(5.161)}\\ \cline{2-6}
&W &{\bf 19.514(0.589})&	33.571(7.859)&{\bf	17.201(0.641)}&	40.142(2.378)\\ \cline{2-6}
%& WMOM& 21.366(0.642)&	38.924(3.759)&	20.193(1.1)	&40.529(1.159)\\ \cline{2-6}
& MVE &20.391(0.715)&	20.348(0.758)&	25.36(2.311)&	26.942(0.988)\\ \cline{2-6}
& MCD &20.296(0.691)&{\bf 	20.12(0.55)}&	25.546(1.801)&26.753(0.864)\\ \cline{2-6}
& M &20.935(0.751)&	45.615(4.473)&	18.898(0.786)	&41.615(0.764)\\ \cline{2-6}
& S& 20.356(0.617)&	20.393(0.693)&	25.707(1.309)	&{\bf 26.624(0.825)}\\ \cline{2-6}
& SD & 20.631(0.67)&	20.61(0.78)&25.672(2.068)&	27.133(1.078)\\ \cline{2-6}
\hline
\end{tabular}}
	\end{table}
	
For example, it has been observed that for the Normal distribution,  the improvement is in the range of 38\%-  50\% 
when training  sets  are {\it mildly} contaminated, whereas the improvement shoots to nearly 72\% - 84\% for {\it hard } contamination in the training   sets. 
Though theoretically the test set is unknown, for the sake of comparison, we also  study  the improvements by using the RGQDA classifiers
when both the training set   and the test  set are contaminated. 
The same scenario prevails in this case, showing the improvement for {\it mild } ({\it hard }) contamination in the range of nearly 17\% - 45.5\% ( 58\% - 66\%). 
For the  t distribution with 3 d.f., the improvement is 15.3\% - 19.6\% ( 45.3\%-72.7\%)  in the case of {\it mild } ({\it hard }) contamination of the  training  set,
whereas  the improvement is 6.9\% - 9.7\% ( 41.6\% - 61.1\%) in the case of {\it mild } ({\it hard }) contamination of both the training  and the test  set,
except for few cases  where the  GQDA classifier still performs reasonably well.
%(???????? 15\% train and test contamination)
%Sometimes the classifier using  GQDA performs better in case of contamination of both the train and test data set which is a rare phenomenon, though not totally absurd. The study reveals that 
However, for the Cauchy distribution the improvement is rather less compared to the Normal distribution and the t distribution with 3 d.f.,
as expected, due to its heavy-tail nature, although the absolute amount of decrease in ME\% 
is also quite significant using some  RGQDA  classifiers based on the estimators having greater robustness.
The improvement  for {\it mild } ({\it hard }) contamination of the training   set has been noted in the range of 15.9\% -19.7\% ( 1.7\% - 44.8\%), 
while the same for the contamination of both the train and the test  set ranges in  $17.3\%-30\%\ \   ( 24.5\%-41.2\%)$.

In terms of the comparisons among  different robust versions of the GQDA classifier proposed in RGQDA, all of them perform similarly well
for weaker (5\%) contamination in all the cases considered here. 
For higher degree of contamination, better performances are observed by the use of strongly robust estimators like MVE, MCD, S or SD.
However, interestingly, for the case of {\it mild } contamination of both the training  and the test  sets, most of the times the  RGQDA classifier based on W or  M estimator
provides the  best performance compared to the other versions.

%\newpage
\subsection{Performances under contamination: four-class problem}
%\subsection{Multi-class}

The empirical results from the four -class simulations are reported in Table \ref{tab Nm} through Table \ref{tab CAm}
for the three distributions discussed so far,  under different types of contaminations. 
Besides the general trend of increasing improvement as the degree of contamination increases, 
the use of different robust statistics lessens the  \% misclassification error when the contamination is {\it hard } compared to the case  when it is {\it mild }.

%\newpage
\begin{table}[!h]
	\centering
	\caption{ Average ME\% (SD) for four-class problems with contaminated Normal distributions, 
		using the GQDA classifier  and its different versions proposed in RGQDA   }\label{tab Nm}
	\scalebox{0.8}{\begin{tabular}{|c|c|c|c|c|c|}
			\hline
			\multicolumn{1}{|c|} {Degree  }& \multicolumn{1}{c|}{  Method } &\multicolumn{4}{c|}{Contamination type}\\ \cline{3-6}
			\multicolumn{1}{|c|} { of}& of &  \multicolumn{1}{c|}{Train {\it mild }}  &\multicolumn{1}{c|}{ Train {\it hard } }& \multicolumn{1}{|c|}{Train and Test {\it mild }} &  \multicolumn{1}{|c|}{Train and Test {\it hard }}\\
			\cline{3-6}
			\multicolumn{1}{|c|}{contamination } & \multicolumn{1}{|c|}{classification} & mean (S.D.) & mean (S.D.) &mean(S.D.) & mean (S.D.) \\
			\hline
			& GQDA & \textbf{12.126}(1.824) &\textbf{26.956}(2.788) &\textbf{10.475}(1.403)  &\textbf{ 31.803}(1.969) \\ \cline{2-6}
			& W & 8.3(0.920) &\textbf{7.566}(1.040 &8.55667(0.995) &12.093(0.967) \\ \cline{2-6}
			& MVE & \textbf{7.696}(0.766) &8.753(1.076) &9.05333(1.036) &\textbf{11.26} (0.959) \\ \cline{2-6}
			5\% & MCD & 7.96(0.982) &13.566(15.704) &8.83167(1.124) &12.726(0.950) \\ \cline{2-6}
			& M & 9.023(0.951) &8.943(0.941) &7.97(0.938) &12.516(1.034) \\ \cline{2-6}
			& S & 8.396(1.104) &10.9(1.240) &9.608(1.427) &13.9(1.095) \\ \cline{2-6}
			& SD & 8.76(1.147) &9.6(9.494) &\textbf{7.898}(0.973) &11.796(0.933) \\ \hline\hline
			& GQDA & \textbf{13.696}(1.701) &\textbf{52.086}(2.169) &\textbf{11.896}(1.501) &\textbf{ 53.376}(2.036) \\ \cline{2-6}
			& W & 8.646(1.017) &9.376(1.005) &7.813(1.085) &\textbf{ 17.263}(0.970) \\ \cline{2-6}
			& MVE & 8.386(0.879) &\textbf{ 7.603}(0.884) &7.891(1.168) &17.72(0.962) \\ \cline{2-6}
			10\% & MCD & 8.78(0.931) &8.676(0.888) &7.831(1.254) &18.393(0.840) \\ \cline{2-6}
			& M & 8.526(1.099) &11.283(1.754) &8.108(1.030) &23.18(1.475) \\ \cline{2-6}
			& S & 8.303(0.864) &8.03(0.859) &8.353(1.304) &17.9(0.992) \\ \cline{2-6}
			& SD & \textbf{8.096}(1.055) &8.74(1.040) &\textbf{ 7.428}(1.066) &18.6(8.186) \\ \hline\hline
			& GQDA & \textbf{13.07}(1.596) &\textbf{44.09}(4.165) &\textbf{13.006}(1.555) &\textbf{47.603}(2.969) \\ \cline{2-6}
			& W & 8.29(0.937) &10.483(0.847) &\textbf{7.42}(1.004) &22.7(1.070) \\ \cline{2-6}
			& MVE & 8.37(1.039) &8.783(0.954) &7.586(1.535) &22.053(1.012) \\ \cline{2-6}
			15\% & MCD & 9.333(1.010) &8.933(0.862) &7.673(1.365) &\textbf{21.503}(0.919) \\ \cline{2-6}
			& M & 9.366(1.292) &49.503(2.918) &9.24(1.188) &49.8(2.596) \\ \cline{2-6}
			& S & 10.093(0.973) &9.123(1.093) &10.466(1.871) &22.46(1.012) \\ \cline{2-6}
			& SD & \textbf{7.61}(0.886) &\textbf{ 8.713}(0.857) &7.523(0.902) &21.55(0.852) \\ \hline\hline
			& GQDA & \textbf{13.42 }(1.381) &\textbf{34.74}(3.061) &\textbf{10.84}(1.607) &\textbf{46.13}(3.573) \\ \cline{2-6}
			& W & 10.173(1.171) &26.59(2.469) &9.03(1.394) &36.76(1.902) \\ \cline{2-6}
			& MVE & 8.813(1.124) &\textbf{8}(1.067) &7.723(0.945) &26.543(0.885) \\ \cline{2-6}
			20\% & MCD & 8.073(1.006) &8.986(1.163) &\textbf{7.613}(1.222) &28.07(1.111) \\ \cline{2-6}
			& M & 11.226(1.368) &45.83(4.500) &9.85(1.211) &50.56(3.315) \\ \cline{2-6}
			& S & \textbf{8.04}(0.899) &9.31(0.984) &7.776(1.302) &\textbf{25.91 }(0.942) \\ \cline{2-6}
			& SD & 8.136(0.994) &10.61(9.342) &8.316(1.605) &27.137(0.899) \\  
			\hline
	\end{tabular}}
\end{table}

In particular, for the Normal distribution with four  classes to classify into, 
the improvement is in the range of $36 \%$ to $42\%$ with {\it mild } contamination 
and in the range of $72\%$ to $86\%$  in the case of {\it hard } contamination in the train set. 
Similar to the previous comparisons, we study the improvement when both the train and the (theoretically unknown) test  set are contaminated.  
Similar to the two-class problems studied before, 
  an improvement in the range of $24\%$ to $43\%$ has been observed in the case of {\it mild} contamination whereas the corresponding range is  $45\%$ to $67\% $ 
for {\it hard } contamination.

%\newpage
\begin{table}[!h]
	\centering
	\caption{ Average ME\% (SD)  in the test data set for four-class problems with  contaminated $t$ distributions with $3$ df, 
		using the GQDA classifier  and its different robust versions proposed in RGQDA   }\label{tab t3m}
	\scalebox{0.8}{\begin{tabular}{|c|c|c|c|c|c|}
			\hline
			\multicolumn{1}{|c|} {Degree  }& \multicolumn{1}{c|}{  Method } &\multicolumn{4}{c|}{Contamination type}\\ \cline{3-6}
			\multicolumn{1}{|c|} { of}& of &  \multicolumn{1}{c|}{Train {\it mild }}  &\multicolumn{1}{c|}{ Train {\it hard } }& \multicolumn{1}{|c|}{Train and Test {\it mild }} &  \multicolumn{1}{|c|}{Train and Test {\it hard }}\\
			\cline{3-6}
			\multicolumn{1}{|c|}{contamination } & \multicolumn{1}{|c|}{classification} & mean (S.D.) & mean (S.D.) &mean(S.D.) & mean (S.D.) \\
			\hline
			
			& GQDA & \textbf{23.453}(2.084) & \textbf{32.573}(2.523) & \textbf{23.417}(2.214) & \textbf{35.75}(2.616) \\ \cline{2-6}
			& W & 17.323(1.176) & 16.69(1.184) & \textbf{16.327}(0.983) & \textbf{20.75}(1.438) \\ \cline{2-6}
			& MVE & 17.097(1.354) & 18.706(1.408) & 19.293(1.363) & 21.08(1.242) \\ \cline{2-6}
			5\% & MCD & 17.91(1.489) & 16.99(1.247) & 16.813(1.365) & 21.77(1.371) \\ \cline{2-6}
			& M & \textbf{16.573}(1.346) & 19.063(1.469) & 16.53(1.451) & 22.317(1.31) \\ \cline{2-6}
			& S & 17.56(1.321) & \textbf{16.207}(1.42919) & 18.21(1.45) & 22.387(1.260) \\ \cline{2-6}
			& SD & 17.31(1.179) & 18.61(1.237) & 17.337(1.277) & 20.84(1.404) \\ \hline\hline
			& GQDA & \textbf{22.76}(2.056) & \textbf{58.343}(5.557) & \textbf{20.227}(1.895) & \textbf{59.403}(3.937) \\ \cline{2-6}
			& W & \textbf{17.687}(1.345) & 18.7(1.605) & \textbf{15.573}(1.031) & 26.23(1.361) \\ \cline{2-6}
			& MVE & 17.907(1.361) & 17.067(1.314) & 16.653(1.708) & \textbf{25.727}(1.261) \\ \cline{2-6}
			10\% & MCD & 17.853(1.392) & \textbf{17.053}(1.236) & 16.613(1.509) & 26.4(1.255) \\ \cline{2-6}
			& M & 18.073(1.261) & 22.697(1.772) & 16.71(1.28) & 28.29(1.602) \\ \cline{2-6}
			& S & 17.873(1.481) & 17.927(1.417) & 18.83333(2.127) & 26.227(1.214) \\ \cline{2-6}
			& SD & 18.113(1.174) & 17.48(1.437) & 16.437(1.75) & 26.133(1.351) \\ \hline\hline
			& GQDA & \textbf{22.233}(1.731) & \textbf{51.72}(2.810) & \textbf{ 23.08}(1.705) & \textbf{56.093}(2.788) \\ \cline{2-6}
			& W & 18.53(1.398) & 22.81(1.608) & 15.657(1.514) & 30.7(1.449) \\ \cline{2-6}
			& MVE & \textbf{17.093}(1.451) & 17.813(1.410) & 18.573(3.200) & 30.63(1.229) \\ \cline{2-6}
			15\% & MCD & 17.117(1.346) & 17.933(1.12738) & 18.56(2.268) & 32.213(8.905) \\ \cline{2-6}
			& M & 19.363(1.513) & 54.277(3.857) & 17.78667(1.826) & 55.843(2.981) \\ \cline{2-6}
			& S & 17.787(1.361) & \textbf{16.787}(1.368) & 18.617(2.089) & \textbf{29.823}(1.416) \\ \cline{2-6}
			& SD & 18.01(1.472) & 18.347(1.167) & \textbf{15.583}(1.338) & 30.203(1.182) \\ \hline\hline
			& GQDA & \textbf{22.64}(2.195) & \textbf{45.877}(4.508) & \textbf{20.923}(1.900) & \textbf{ 52.31 }(4.467) \\ \cline{2-6}
			& W & 21.977(1.675) & 51.277(3.472) & 18.267(1.775) & 52.69(3.250) \\ \cline{2-6}
			& MVE &\textbf{ 17.84}(5.531) & 17.717(1.344) & 17.307(3.126) & \textbf{31.65}(2.226) \\ \cline{2-6}
			20\% & MCD & 18.153(1.292) & 16.477(1.72155) & 19.283(3.393) & 33.207(0.996) \\ \cline{2-6}
			& M & 21.337(1.783) & 56.173(9.229) & 18.323(1.748) & 52.743(5.910) \\ \cline{2-6}
			& S & 18.863(1.077) & \textbf{15.643}(1.469) & \textbf{15.9 }(2.611) & 34.437(1.329) \\ \cline{2-6}
			& SD & 18.2(1.305) & 17.747(1.433) & 16.317(2.218) & 33.423(1.246) \\ \hline
	\end{tabular}}
\end{table}

%\newpage
\begin{table}[!h]
	\centering
	\caption{Average ME\% (SD) for four-class problems with contaminated Cauchy distributions, 
		using the GQDA classifier and its  different robust versions proposed in  RGQDA   }\label{tab CAm}
	\scalebox{0.8}{\begin{tabular}{|c|c|c|c|c|c|}
			\hline
			\multicolumn{1}{|c|} {Degree  }& \multicolumn{1}{c|}{  Method } &\multicolumn{4}{c|}{Contamination type}\\ \cline{3-6}
			\multicolumn{1}{|c|} { of}& of &  \multicolumn{1}{c|}{Train {\it mild }}  &\multicolumn{1}{c|}{ Train {\it hard } }& \multicolumn{1}{|c|}{Train and Test {\it mild }} &  \multicolumn{1}{|c|}{Train and Test {\it hard }}\\
			\cline{3-6}
			\multicolumn{1}{|c|}{contamination } & \multicolumn{1}{|c|}{classification} & mean (S.D.) & mean (S.D.) &mean(S.D.) & mean (S.D.) \\
			\hline
			& GQDA & \textbf{45.193}(3.691) & \textbf{ 49.77}(3.960) & \textbf{ 57.923 }(6.113) & \textbf{52.546}(5.897) \\ \cline{2-6}
			& W & 30.353(1.835) & \textbf{29.626}(1.507) & 32.93(1.874) & \textbf{32.58 }(1.496) \\ \cline{2-6}
			& MVE & \textbf{29.436}(1.775) & 30.31(1.676) & 33.843(1.612) & 32.766(1.898) \\ \cline{2-6}
			5\% & MCD & 30.37(1.570) & 28.55(1.738) & 33.446(1.448) & 34.03(1.623) \\ \cline{2-6}
			& M & 30.17(1.610) & 32.773(1.591) & 34.15(1.525) & 34.95(1.558) \\ \cline{2-6}
			& S & 31.05(1.325) & 29.93(1.458) & 32.693(1.416) & 32.76(1.452) \\ \cline{2-6}
			& SD & \textbf{29.437}(1.574) & 30.943(1.477) & \textbf{32.65}(6.361) & 33.057(2.588) \\ \hline\hline
			& GQDA & \textbf{43.85}(5.297) & \textbf{75.193}(7.155) & \textbf{43.507}(3.640) & \textbf{ 75.617}(8.686) \\ \cline{2-6}
			& W & 31.107(1.964) & 34.82(6.074) & 28.847(1.706) & 38.037(1.575) \\ \cline{2-6}
			& MVE & 30.133(1.31406) & \textbf{29.5}(1.479) & 30.063(2.59) & \textbf{36.07}(1.457) \\ \cline{2-6}
			10\% & MCD & \textbf{29.013}(1.780) & 32.07(1.620) & 29.653(1.804) & 37.34(5.611) \\ \cline{2-6}
			& M & 31.543(1.961) & 35.537(2.177) & \textbf{27.357}(2.112) & 38.343(3.868) \\ \cline{2-6}
			& S & 31.457(1.294) & 29.92(1.632) & 30.143(2.390) & 38.527(5.429) \\ \cline{2-6}
			& SD & 31.376(1.702) & 30.053(1.369) & 30.2(2.480) & 36.533(1.239) \\ \hline\hline
			& GQDA & \textbf{43.307}(5.172) & \textbf{74.34 }(7.800) & \textbf{40.997 }(4.341) & \textbf{72.613}(9.159) \\ \cline{2-6}
			& W & 32.49(1.798) & 51.917(5.687) & 28.32(2.142) & 47.83(2.358) \\ \cline{2-6}
			& MVE & 31.003(1.469) & 32.607(1.546) & 31.41(2.67) & \textbf{39.993}(1.335) \\ \cline{2-6}
			15\% & MCD & \textbf{28.987}(1.564) & 31.24(1.960) & \textbf{27.98}(2.268) & 40.34(5.140) \\ \cline{2-6}
			& M & 30.897(1.820) & 65.003(7.229) & 29.183(1.849) & 62.563(4.898) \\ \cline{2-6}
			& S & 31.52(1.772) & 31.903(6.414) & 30.467(2.814) & 46.147(11.908) \\ \cline{2-6}
			& SD & 32.093(2.635) & \textbf{30.963}(1.508) & 29.817(2.363) & 40.74(1.782) \\ \hline\hline
			& GQDA & \textbf{40.2 }(6.182) & \textbf{ 65.187}(10.238) & \textbf{42.22}(5.169) & \textbf{75.097}(7.615) \\ \cline{2-6}
			& W & 30.037(1.821) & 66.287(7.210) & 27.737(1.678) & 65.967(7.313) \\ \cline{2-6}
			& MVE & 32.05(1.878) & 30.933(6.622) & 34.743(4.19) & 44.95(1.958) \\ \cline{2-6}
			20\% & MCD & 31.37(4.010) & 30.317(1.553) & 35.903(3.093) & 41.9(1.700) \\ \cline{2-6}
			& M & 31.38(2.043) & 64.99(11.06) & 27.963(1.915) & 62.087(6.732) \\ \cline{2-6}
			& S & \textbf{30.007}(1.981) & \textbf{30.28}(6.652) & 33.433(3.166) & \textbf{41.88}(1.619) \\ \cline{2-6}
			& SD & 31.883(1.323) & 31.38(1.668) & \textbf{27.163}(2.390) & 55(2.659) \\\hline
	\end{tabular}}
\end{table}

For the $t$ distribution with 3 d.f. with  four classes to classify into, 
the improvement is in the range of $21\%$ to $29\%$ with {\it mild } contamination, 
and in the range of  $50\%$ to $71\%$  in the case of {\it hard } contamination in the train set. 
Similar to the previous comparisons, we study the improvement when both the train and the (theoretically unknown) test set are contaminated.  
Once again, the use of different RGQDA classifiers results into  an improvement in the range of $22\%$ to $32\%$ for {\it mild} contamination, and in the range of  $39\%$ to $57\%$ for {\it hard } contamination. In summary, the justification for the use of different robust versions of the GQDA classifier proposed in RGQDA is emphatically evident from the simulation studies.

%\cleardoublepage

%\cleardoublepage
\newpage
%\bigskip\bigskip
\section{Real Data Applications}
\label{SEC:real_data}

To get a sense of the performances of  different  RGQDA classifiers  in the real data scenario,  we have also applied the proposed methods to classify  several real data sets 
obtained from the UCI Machine Learning Repository \citep{Dua/Graff:2019}. To our expectation, all the findings are seen to support 
the claim that different  RGQDA classifiers outperform the GQDA classifier.
However, again for brevity, we present the results for three such data sets 
from wider application ranges, namely astrophysics, business and biomedical domains.

\subsection{Data sets}

Let us first present a brief description of each of the three data sets along with the feature lists that we have used in our illustration.

\begin{itemize}
	\item  \textbf{Ionosphere Data:} These radar data, collected by a system consisting of a phased array of 16 high-frequency antennas 
	in Goose Bay, Labrador, consist of 351 observations (radar returns) classified into two classes, along with 34 associated features.
	The two classes are referred to as \textit{Good} and \textit{Bad},  respectively, depending on  
	whether the corresponding signal is returned from the ionosphere indicating some structure there 
	or it is passed through the ionosphere. 
	An auto-correlation function is used  to process the received signals 
	with input as the time of a pulse and 17 pulse numbers,
	which results into 34 covariates corresponding to the real and imaginary parts of the complex electromagnetic signal output
	from each of the 17 pulse numbers. 
	However, since the variation in the first two variables are observed to be zero, 
	the remaining 32 variables (numbered 3-34) are used to classify the objects into the above-mentioned two classes. 
%	The number of observations available in the data is $351$.

	\item \textbf{Statlog ACA Data:} These data correspond to Australian credit card applications (suitably modified to maintain confidentiality).
	There are 690 observations belonging to either of the two classes recorded in an (anonymous) class attribute, 
	along with  8 categorical features and 6 numerical features. 
	We consider the 5 continuous variables (except the variable A10) to examine our RGQDA,
	since the \textit{SD} estimator becomes computationally instable if A10 is included.

	\item  \textbf{New Thyroid Data:} This data set, donated by Stefan Aeber{\it hard } from Garavan Institute in Sydney, Australia,
	contains information about 215 patients classified into three classes corresponding to euthyroidism, hypothyroidism or
	hyperthyroidism thyroid. The purpose is to form a predictive model to classify patient's thyroid type into the above three classes
	based on 5 features which are 5 laboratory tests, namely T3-resin uptake test (in percentage), 
	total Serum thyroxin as measured by the isotopic     displacement method, 
	total serum triiodothyronine as measured by radioimmuno assay, 
	basal thyroid-stimulating hormone (TSH) as measured by radioimmuno assay,
	and maximal absolute difference of TSH value after injection of  200 micro grams of thyrotropin-releasing hormone as compared
    to the basal value. We have used all these five continuous feature variables to perform the proposed RGQDA. 
\end{itemize}

\begin{figure}[!h]
	\centering
	\includegraphics[scale=0.45]{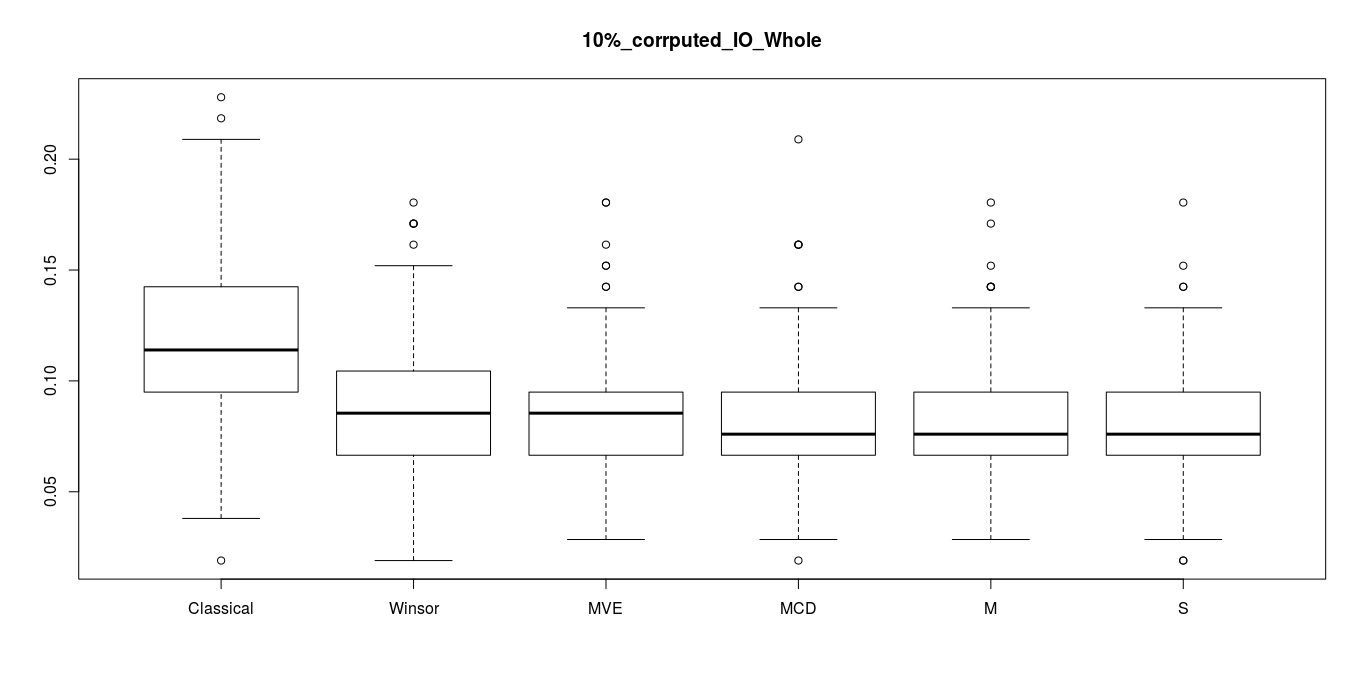}
	\caption{ Boxplots of ME\%s using  the  GQDA classifier and  different RGQDA classifiers  for the Ionosphere data}
	\label{FIG:1}
\end{figure}

\begin{figure}[!h]
	\centering
	\includegraphics[width=165mm,scale=0.6]{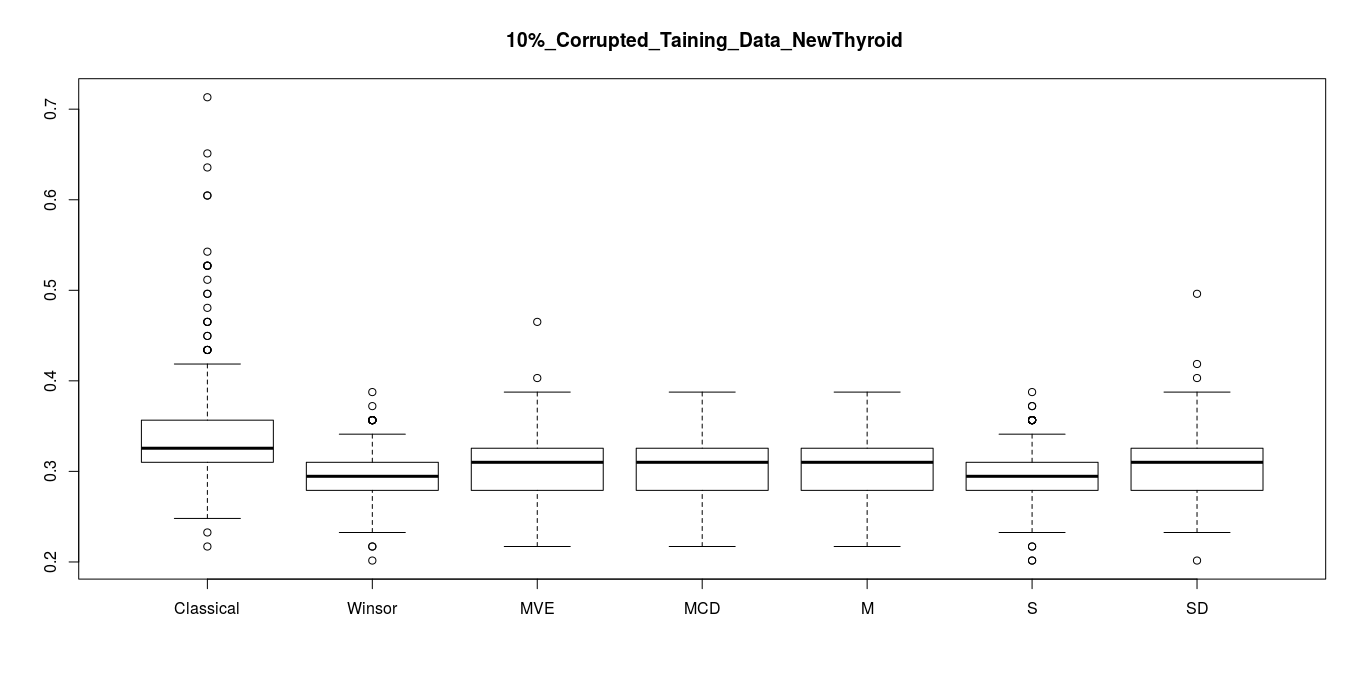}
	\caption{ Boxplots of ME\%s using  the  GQDA classifier and  different RGQDA classifiers for the New Thyroid data}
	\label{FIG:3}
\end{figure}

For each of these three datasets, all available observations are randomly partitioned into the training set and the test  set 
of sizes approximately $70\%$ and  $30\%$, respectively, of the entire data set. 
Further, for robustness illustrations, $10\%$ of the training  set has been forcefully misclassified to act like {\it outliers}.
Then, we have applied different  proposed RGQDA classifiers  along with the  GQDA classifier for the classification of the modified data set.
The thresholds  obtained from the contaminated training  set  are then used to compute 
the \% misclassification errors in the corresponding test set.
This process is repeated 500 times with different random partition in each of the replications, 
and the boxplots of the resulting ME\% s  are presented for each of the proposed  RGQDA classifier  and the  GQDA classifier 
in  Figure \ref{FIG:1} through  Figure \ref{FIG:3},  respectively, for the three above-mentioned datasets.

%\begin{center}
%    \centering
%    \includegraphics[width=165mm,scale=0.6]{1vsRest.png}
%\end{center}
%
%\begin{center}
%    \centering
%    \includegraphics[width=165mm,scale=0.6]{2vsRest.png}
%\end{center}
%
%\begin{center}
%    \centering
%    \includegraphics[width=165mm,scale=0.6]{3vsRest.png}
%\end{center}
%
%\begin{center}
%    \centering
%    \includegraphics[width=165mm,scale=0.6]{3vs1.png}
%\end{center}
%
%\begin{center}
%    \centering
%    \includegraphics[width=165mm,scale=0.6]{2vs3.png}
%\end{center}
%
%\begin{center}
%    \centering
%    \includegraphics[width=165mm,scale=0.6]{2vs1.png}
%\end{center}
%
%     

\begin{figure}[!h]
	\centering
	\includegraphics[scale=0.45]{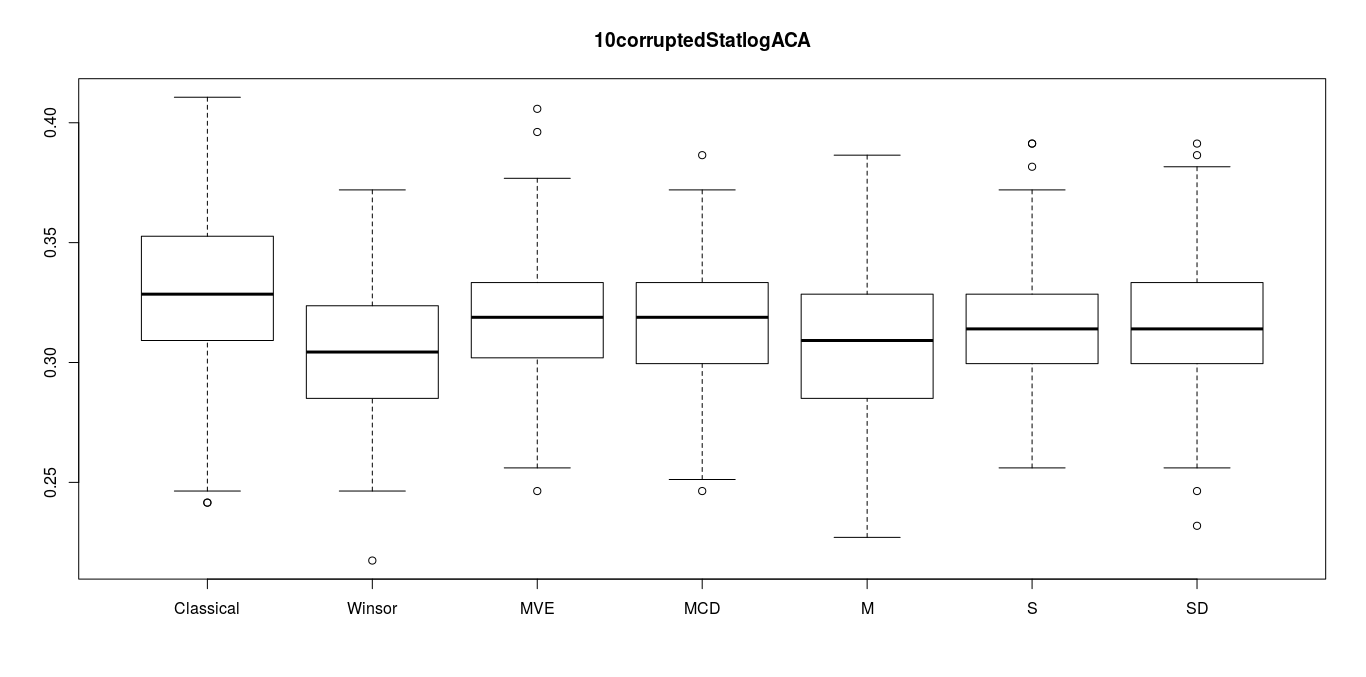}
	\caption{Boxplots of ME\%s using  the  GQDA classifier and  different RGQDA classifiers  for the Statlog ACA data}
\end{figure}

\subsection{Results}

For all the three above-mentioned data sets (along with several others not reported here for brevity),
the advantages of using different  RGQDA classifiers over the  GQDA classifier is emphatically clear in the presence of contamination.
In particular, for the ionosphere data, the box-plots of the ME\%s  for each of  the RGQDA classifiers is way below than 
the one  obtained using the the GQDA classifier. This finding  is also consistent with what has been observed in our simulation  
studies, indicating  better performances of the RGQDA classifiers.
For the multi class classification in the New Thyroid data as well, 
it has been resoundingly reinforced that the RGQDA method works much better than  the GQDA classifier 
with significantly lesser spread of the box-plots of the \%  misclassification errors.
Among the different robust versions proposed in RGQDA, the classifiers based on W and S estimators perform the best for the New Thyroid data set,
whereas all other classifiers  using  MCD, M and S estimators exhibit a similar better performance  for the Ionosphere data. For the Statlog ACA data,
the results are comparatively more sparsed, with the classifier based on  W estimator providing the best performance in RGQDA. Overall, the justification of the use of different  RGQDA classifiers  for the purpose of classification is profoundly demonstrated in all the three data sets.

\section{Conclusion and discussions}
 \label{SEC:conclusion}

In this paper, for the multivariate classification problem, in the likely presence of outliers in the data set, an attempt has been made to study the performance of  the GQDA classifier, 
which is  a generalization of the  QDA classifier  and the MMD  classifier, proposed by \cite{Bose/etc:2015} and suggest a better strategy. 
 Through the simulation studies of some elliptically symmetric distributions as well  as working with some real data sets, it has been shown 
that the GQDA classifier performs miserably when the data set is contaminated. 
Hence to propose a better alternative,  an investigation has  been made to replace the estimators used in  the GQDA classifier 
by different robust versions of the classical estimator of  the mean  vector and  the dispersion   matrix 
available in the literature and the  new procedure has been termed as RGQDA.  
An overall commendable performance of different  RGQDA classifiers has been transpired so far, in all the examples we tried with, 
be it simulation or real life data set, varying the degree of improvement across the different robust versions adopted in RGQDA. 
Thus a comparative study of the performances of different RGQDA classifiers, using  different robust versions of the classical estimator 
in the presence of different degree and the type of contamination  suggests  the choice of the particular robust estimator to be adopted, 
to have a reasonably  good  classification, in terms of reducing the \% misclassification error. It is also interesting to note that even when there is no contamination in the data set, the proposed  RGQDA classifiers outperform the GQDA classifier in classifying the underlying populations having distributions with heavier tail than the Normal distribution.
The  performance of different RGQDA classifiers needs to be looked  into  when we face the  dimensionality reduction issue, 
in case of high-dimensional set-ups with large $p$-small $n$  problem. 
Future investigation  in this direction will be reported in a separate  article.  

\bigskip\noindent
\textbf{Acknowledgment:} 
The research of the first author (AG) is partially supported by the INSPIRE Faculty Research Grant from Department of Science and Technology, 
Government of India.


\begin{thebibliography}{3}
	%
%\bibitem[\protect\citeauthoryear{Basu, Basu and Jones}{Basu et~al.}{2006}]{Basu/etc:2006}
%Basu, S., Basu, A., and Jones, M. C. (2006). 
%Robust and efficient parametric estimation for censored survival data. 
%{\em Ann. Inst. Stat. Math.}, {\bf 58(2)}, 341--355.

%\bibitem[\protect\citeauthoryear{Basu, Shioya and Park}{Basu et~al.}{2011}]{Basu/etc:2011}
%Basu, A., Shioya, H. and Park, C. (2011).
%\newblock {\it Statistical Inference: The Minimum Distance Approach}.
%\newblock Chapman \& Hall/CRC. Boca Raton, Florida. 
%

	
\bibitem[\protect\citeauthoryear{Bickel}{Bickel}{1965}]{Bickel:1965}
Bickel, P. J. (1965). 
On some robust estimates of location. 
{\em Ann. Math. Statist.}, 36, 847--858.


\bibitem[\protect\citeauthoryear{Bose et al.}{Bose et al.}{2015}]{Bose/etc:2015}
Bose, S., Pal, A.;  SahaRay, R and Nayak, J. (2015). 
Generalised Quadratic Discriminant Analysis. 
{\em  Pattern Recognit.}, 48,  2676--2684.


\bibitem[\protect\citeauthoryear{Butler}{Butler et al.}{1993}]{Butler/etc:1993}
Butler, R. W.; P. L. Davies and M. Jhun (1993). 
Asymptotics for the minimum covariance determinant estimator. 
{\em Ann. Statist.}, 21, 1385--1400.


\bibitem[\protect\citeauthoryear{Chork and Rousseeuw}{Chork and Rousseeuw}{1992}]{Chork/Rousseeuw:1992}
Chork, C. Y. and Rousseeuw, P. J. (1992). 
Integrating a high-breakdown option into discriminant analysis in exploration geochemicals. 
{\em J. Geoch. Explorat.}, 43, 191--203. 


\bibitem[\protect\citeauthoryear{Croux and Dehon}{Croux and Dehon}{2001}]{Croux/Dehon:2001}
Croux, C. and Dehon, C. (2001). 
Robust linear discriminat analysis using S-estimators. 
{\em Canad. J. Statist.}, 29, 473--492.


\bibitem[\protect\citeauthoryear{Croux and Haesbroec}{Croux and Haesbroec}{1999}]{Croux/Haesbroec:1999}
Croux, C. and G. Haesbroeck (1999). 
Influence function and efficiency of the minimum covariance determinant scatter matrix estimator. 
{\em J. Multivariate Anal.}, 71, 161--190.


\bibitem[\protect\citeauthoryear{Daqi}{Daqi et al.}{2014}]{Daqi/etc:2014}
Daqi, G., Jun, D., and Changming, Z. (2014). 
Integrated Fisher linear discriminants: An empirical study. 
\textit{Pattern Recognit.}, 47(2), 789-805.

	
\bibitem[\protect\citeauthoryear{Davies}{Davies}{1987}]{Davies:1987}
Davies, P. L.  (1987). 
Asymptotic behavior of S-estimators of multivariate location parameters and the dispersion  matrix matrices. 
{\em Ann. Statist.}, 15, 1269--1292.


\bibitem[\protect\citeauthoryear{Davies}{Davies}{1992}]{Davies:1992a}
Davies, P. L.  (1992a). 
The asymptotics of Rousseeuw's minimum volume ellipsoid estimator. 
{\em Ann. Statist.}, 20, 1828--1843.


\bibitem[\protect\citeauthoryear{Donoho}{Donoho}{1982}]{Donoho:1982}
Donoho, D. L.  (1982). 
\textit{Breakdown properties of multiavraite location estiamtors}.
Ph.D. thesis, Harvard University.


\bibitem[\protect\citeauthoryear{Dua, D. and Graff}{Dua and Graff}{2019}]{Dua/Graff:2019}
Dua, D. and Graff, C. (2019). 
\textit{UCI Machine Learning Repository [http://archive.ics.uci.edu/ml]}. 
Irvine, CA: University of California, School of Information and Computer Science.


\bibitem[\protect\citeauthoryear{Haddad}{Haddad et~al.}{2013}]{Haddad/etc:2013}
Haddad, F. S., Syed‐Yahaya, S. S., and Alfaro, J. L. (2013). 
Alternative Hotelling's T2 Charts using Winsorized Modified One‐Step M‐estimator. 
{\em Quality Reliab. Eng. Int.}, 29(4), 583-593.

	
\bibitem[\protect\citeauthoryear{Hampel, Ronchetti, Rousseeuw, and  Stahel}{Hampel et~al.}{1986}]{Hampel/etc:1986}
Hampel, F.~R., E.~Ronchetti, P.~J. Rousseeuw, and W.~Stahel (1986).
\newblock {\em Robust Statistics: The Approach Based on Influence Functions}.
\newblock New York, USA: John Wiley \& Sons.


\bibitem[\protect\citeauthoryear{Hua}{Hua et al.}{2005}]{Hua/etc:2005}
Hua, J., Xiong, Z., and Dougherty, E. R. (2005). 
Determination of the optimal number of features for quadratic discriminant analysis via the normal approximation to the discriminant distribution. 
\textit{Pattern Recognit.}, 38(3), 403-421.




\bibitem[\protect\citeauthoryear{Huber}{Huber}{1981}]{Huber:1981}
Huber, P. J.  (1981). 
\textit{Robust Statistics}. 
John Wiley \& Sons, New York.



\bibitem[\protect\citeauthoryear{Hubert, van Driessen}{Hubert and van Driessen}{2004}]{Hubert/vanDriessen:2004}
Hubert, M.; P. and Driessen,K. van (2004). 
Fast and robust discriminant analysis. 
{\em Comput. Statist. Data. Anal.}, 45(2), 301--320.


\bibitem[\protect\citeauthoryear{Hubert, Rousseeuw, and  Stahel}{Hubert et~al.}{2012}]{Hubert/etc:2012}
Hubert, M.; P. Rousseeuw and T. Verdonck (2012). 
A deterministic algorithm for robust location and scatter. 
{\em Journal of Computational and Graphical Statistics}, 21(3), 618--637.


\bibitem[\protect\citeauthoryear{Kim et al.}{Kim et al.}{2006}]{Kim/etc:2006}
Kim, S-J.; Magnani, A. and Boyd, S.P ( 2006).
Robust Fisher Discriminant Ananlysis.
{\em Adv. Neural Info. Process. Sys.}, 18, 659--666.



\bibitem[\protect\citeauthoryear{Lopuhaa}{Lopuhaa}{1989}]{Lopuhaa:1989}
Lopuhaa, H. P.  (1989). 
On the relation between S-estimators and M-estimators of multivariate location and covariance. 
{\em Ann. Statist.},  17, 1662--1683.



\bibitem[\protect\citeauthoryear{Maronna}{Maronna}{1976}]{Maronna:1976}
Maronna, R. A. (1976). 
Robust M-estimators of multivariate location and scatter. 
{\em Ann. Statist.}, 4(1), 51--67.


\bibitem[\protect\citeauthoryear{Maronna and Yohai}{Maronna and Yohai}{1995}]{Maronna/Yohai:1995}
Maronna, R. A., and V. J. Yohai (1995). 
The behavior of the Stahel-Donoho  robust multivariate estimator. 
{\em J. Amer. Statist. Assoc.}, 90, 330--341.


\bibitem[\protect\citeauthoryear{Zollanvari}{Na et al.}{2010}]{Na/etc:2010}
Na, J. H., Park, M. S., and Choi, J. Y. (2010). 
Linear boundary discriminant analysis. 
\textit{Pattern Recognit.}, 43(3), 929--936.


\bibitem[\protect\citeauthoryear{Park 2}{Park and Park}{2008}]{Park/Park:2008}
Park, C. H., and Park, H. (2008). 
A comparison of generalized linear discriminant analysis algorithms. 
\textit{Pattern Recognit.}, 41(3), 1083--1097.


\bibitem[\protect\citeauthoryear{Randles et al.}{Randles et al.}{1978}]{Randles/etc:1978}
Randles, R. H. ; Broffitt, J. D.;  Ramsberg, J. S. and Hogg, R. V.(1978). 
Generalized linear and quadratitc discriminant functions using robust estimators.. 
{\em J. Amer. Statist. Assoc.}, 73, 564--568.



\bibitem[\protect\citeauthoryear{Rousseeuw}{Rousseeuw}{1985}]{Rousseeuw:1985}
Rousseeuw, P. J.  (1985). 
Multivariate estimation with high breakdown point.
In {\em Mathematical Statistics and Applications} (Eds. W. Grossmann, G. Pflug, I. Vincze and W. Wertz). 
Reidel. 283--297.


\bibitem[\protect\citeauthoryear{Rousseeuw and van Driessen}{Rousseeuw and van Driessen}{1999}]{Rousseeuw/vanDriessen:1999}
Rousseeuw, P. J. and van Driessen, K. (1999). 
A fast algorithm for the minimum covariance determinant estimator. 
{\em Technometrics}, 41, 212--223.


\bibitem[\protect\citeauthoryear{Stahel}{Stahel}{1981}]{Stahel:1981}
Stahel, W. A.  (1981). 
\textit{Robust estimation: Infinitesimal optimality and covariance matrix estimators}. 
Ph.D. thesis, ETH, Zurich.

\bibitem[\protect\citeauthoryear{Suzuki, M., and Itoh}{Suzuki and Itoh}{2010}]{Suzuki/Itoh:2010}
Suzuki, M., and Itoh, A. (2010). 
Reduction of processing time for optimal and quadratic discriminant analyses. 
\textit{Pattern Recognit.}, 43(9), 3144-3150.


\bibitem[\protect\citeauthoryear{Todorov et al. }{Todorov et al.}{1990}]{Todorov/etc:1990}
 Todorov, V.; Neykov, N. and Neytchev, P. (1990). 
Robust selection of variables in the discriminant analysis based on MVE and MCD estimators. 
{\em Proceed. Computat. Statist.}, COMPSTAT, Physica Verlag, Heidelberg.


\bibitem[\protect\citeauthoryear{Todorov et al. }{Todorov et al.}{1994}]{Todorov/etc:1994}
 Todorov, V.; Neykov, N. and Neytchev, P. (1994). 
Robust two-group discrimination by bounded influence regression. 
{\em Comput. Statist. Data. Anal.}, 17, 289--302.


\bibitem[\protect\citeauthoryear{Zollanvari}{Ye et al.}{2017}]{Ye/etc:2017}
Ye, H., Li, Y., Chen, C., and Zhang, Z. (2017). 
Fast Fisher discriminant analysis with randomized algorithms. 
\textit{Pattern Recognit.}, 72, 82--92.


\bibitem[\protect\citeauthoryear{Zollanvari}{Wang et al.}{2008}]{Wang/etc:2008}
Wang, J., Plataniotis, K. N., Lu, J., and Venetsanopoulos, A. N. (2008). 
Kernel quadratic discriminant analysis for small sample size problem. 
\textit{Pattern Recognit.}, 41(5), 1528--1538.


%\bibitem[\protect\citeauthoryear{Wilcox, R. R., and Keselman}{Wilcox and Keselman}{2003}]{Wilcox/Keselman:2003}
%Wilcox, R. R., and Keselman, H. J. (2003). 
%Repeated measures one‐way ANOVA based on a modified one‐step M‐estimator. 
%British Journal of Mathematical and Statistical Psychology, 56(1), 15-25.


\bibitem[\protect\citeauthoryear{Zollanvari}{Zollanvari et al.}{2013}]{Zollanvari/etc:2013}
Zollanvari, A., Hua, J., and Dougherty, E. R. (2013). 
Analytical study of performance of linear discriminant analysis in stochastic settings. 
\textit{Pattern Recognit.}, 46(11), 3017-3029.



\bibitem[\protect\citeauthoryear{Zuo, Y. J., and Cui}{Zuo and Cui}{2004}]{Zuo/Cui:2004}
Zuo, Y. J., and Cui, H. J. (2004). 
Statistical depth functions and some applications. 
{\em Adv. Math.}, 33(1), 1--25.	

	


%Sign consistency for the linear programming discriminant rule
%
%
%Discriminant component analysis via distance correlation maximization
%
%
%Supervised discrete cross-modal hashing based on kernel discriminant analysis
%
%
%Enhanced Grassmann discriminant analysis with randomized time warping for motion recognition
%
%
%Towards fast and kernelized orthogonal discriminant analysis on person re-identification
%
%
%Multi-view common component discriminant analysis for cross-view classification
%
%
%Fast semi-supervised discriminant analysis for binary classification of large data sets

%\bibitem[\protect\citeauthoryear{Zuo, Y. J., and Cui}{Zuo and Cui}{2004}]{Zuo/Cui:20041}
%Regularized max-min linear discriminant analysis
%
%\bibitem[\protect\citeauthoryear{Zuo, Y. J., and Cui}{Zuo and Cui}{2004}]{Zuo/Cui:20041}
%Regularized orthogonal linear discriminant analysis
%
%\bibitem[\protect\citeauthoryear{Zuo, Y. J., and Cui}{Zuo and Cui}{2004}]{Zuo/Cui:20041}
%L1-norm and maximum margin criterion based discriminant locality preserving projections via trace Lasso

%%%%%%%%%%%%%%%%%%%%%%%%%%%%%%%%%%%%%%%%%%%%%%%%%%%



%\bibitem[\protect\citeauthoryear{Raudys, S., and Duin}{Raudys and Duin}{1998}]{Raudys/Duin:1998}
%Raudys, S., and Duin, R. P. (1998). 
%Expected classification error of the Fisher linear classifier with pseudo-inverse covariance matrix. Pattern recognition letters, 19(5-6), 385-392.















\end{thebibliography}
\end{document}